\documentclass[a4paper,11pt]{article}
\pdfoutput=1
\usepackage{jheppub}

\usepackage{epsf}
\usepackage{epsfig}
\usepackage{subfigure}
\usepackage{mathtools}
\usepackage{hhline}
\usepackage{float}
\usepackage{multirow}
\usepackage{nicefrac}
\usepackage{epstopdf}
\usepackage{url}
\usepackage{multicol}
\usepackage{array}
\usepackage[normalem]{ulem}
\usepackage{slashed}

\allowdisplaybreaks

\newcommand{\be}{\begin{eqnarray}}
\newcommand{\ee}{\end{eqnarray}}

\newcommand{\ba}{\begin{array}}
\newcommand{\ea}{\end{array}}
\newcommand{\bee}{\begin{equation}\ba{c}}
\newcommand{\eee}{\ea\end{equation}}

\newcommand{\bi}{\begin{itemize}}
\newcommand{\ei}{\end{itemize}}

\toccontinuoustrue
\DeclareUnicodeCharacter{2212}{\textendash}

\title{Tau-jet signatures of vectorlike quark decays to heavy charged and neutral Higgs bosons }

\author{Radovan Dermisek$^1$,}
\author{Enrico Lunghi$^1$}
\author{Navin McGinnis$^{2,3}$}
\author{Seodong Shin$^{4}$}
\affiliation{
$^1$Physics Department, Indiana University, Bloomington, IN 47405, USA \\
$^2$TRIUMF, 4004 Westbrook Mall, Vancouver, BC, Canada V6T 2A3\\
$^3$High Energy Physics Division, Argonne National Laboratory, Lemont, IL 60439, USA\\
$^4$Department of Physics, Jeonbuk National University, Jeonju, Jeonbuk 54896, Korea\\ 
}
\emailAdd{dermisek@indiana.edu} 
\emailAdd{elunghi@indiana.edu} 
\emailAdd{nmcginnis@triumf.ca}
\emailAdd{sshin@jbnu.ac.kr}

\abstract{
We study $4b+2\tau$ and $4b+1\tau$ signatures of heavy neutral and charged Higgs bosons originating from cascade decays of pair-produced new quarks. Decays of vectorlike quarks through heavy Higgses can easily dominate in the two Higgs doublet model of type-II, and the studied signatures are common to many possible decay chains. 
 We design search strategies for these final states and discuss the mass ranges of heavy Higgs bosons and new quarks that can be explored at the Large Hadron Collider as functions of branching ratios in a model independent way. We further combine the results with a similar study focusing on decays which lead to a $6b$ final state and interpret the sensitivity to charged and neutral Higgs bosons and vectorlike quarks in the type-II two Higgs doublet model. We find that the LHC reach for their masses extends to well above 2 TeV  in the case of an SU(2) doublet quark and to at least 1.8 TeV for a bottom-like SU(2) singlet quark in the whole range of $\tan\beta$ between 1 and 50.
}

\begin{document} 
\maketitle
\flushbottom

\section{Introduction}
\label{sec:intro}

 In two Higgs doublet models (2HDMs) with vectorlike quarks, the decays of vectorlike quarks are often dominated by cascade decays through charged or neutral Higgs bosons leading to signatures with 6 standard model (SM) fermions~\cite{Dermisek:2019vkc}. The heavy Higgs bosons are effectively pair produced with QCD-size cross sections, while the final states with a large number of 3rd generation fermions, resulting from their dominant decay modes, have very small irreducible SM background. This presents a great  opportunity for the Large Hadron Collider (LHC). However, depending on hierarchies in  masses of new quarks and Higgs bosons and other model parameters, many decay chains are possible. Moreover, the rates for several of them are often comparable, diluting individual final states. Thus designing optimal searches for specific possibilities is not practical, and focusing on the common signature of a whole class of decay chains might be more beneficial.

In ref.~\cite{Dermisek:2020gbr} we investigated final states with combinations of 6 top and bottom quarks in an extension of the type-II 2HDM. In that case, since the top quark also decays into a bottom quark, a common signature for many possible decay chains is 6 bottom quarks. Although the search strategies were tailored to the final state consisting of exactly 6 bottom quarks, they were found to be very effective for all the other processes. Even with current data, new quarks and Higgs bosons with masses of about 1.5 TeV could be seen irrespective of $\tan \beta$ which is striking in comparison with the reach for heavy Higgs bosons in 2HDMs without vectorlike matter. On the other hand, the current limits for heavy neutral CP-even and CP-odd Higgs bosons, $H$ and $A$, are only about 400 GeV for medium  $\tan \beta$~\cite{Aad:2020zxo, Sirunyan:2018zut} and are significantly weaker for the charged Higgs boson, $H^\pm$~\cite{Aaboud:2018gjj, Sirunyan:2019hkq, Aaboud:2018cwk, Sirunyan:2019arl, ATLAS:2020jqj}.

\begin{figure}[t]
\centering
\includegraphics[scale=0.55]{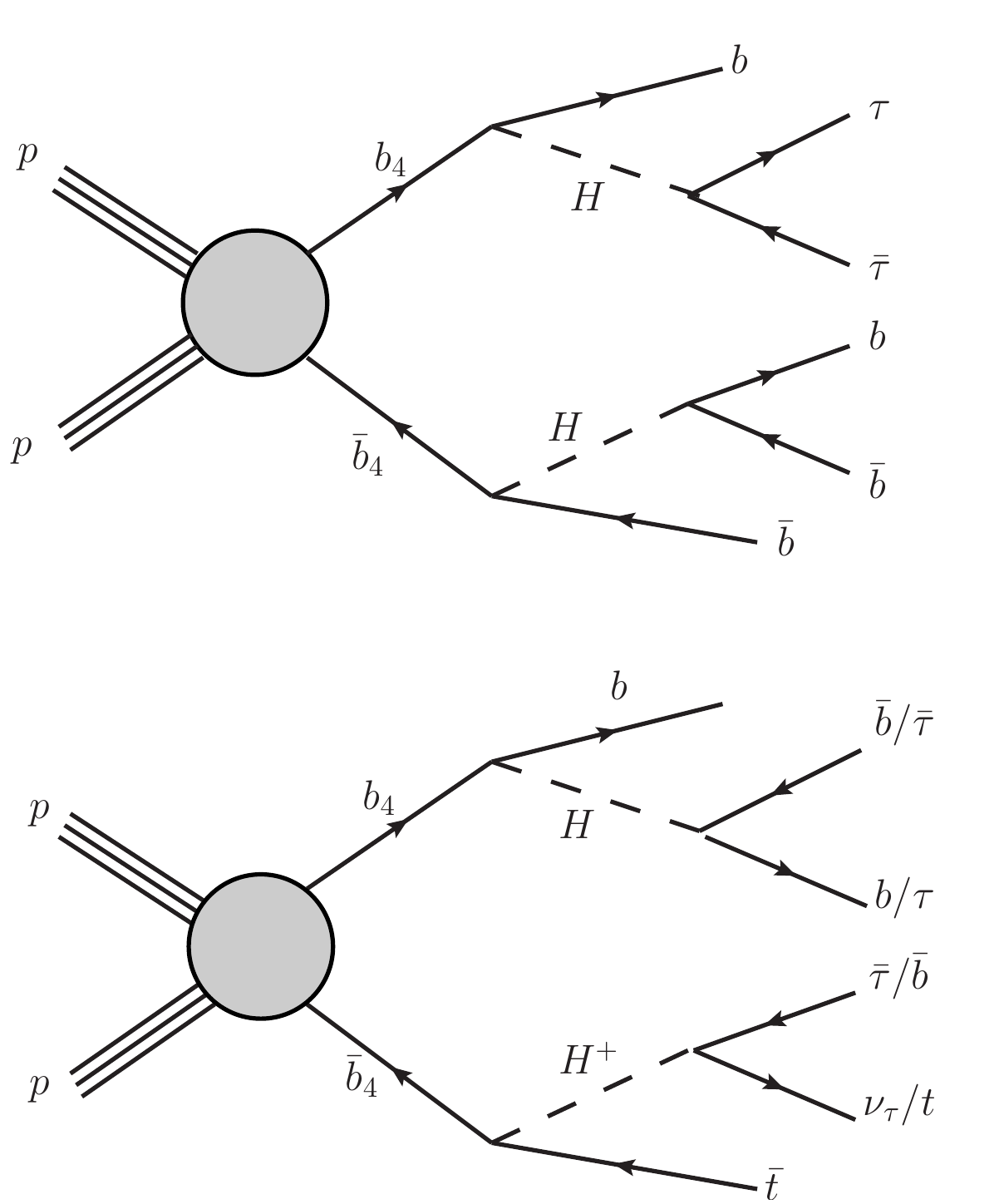}~~~~~~~~~
\includegraphics[scale=0.55]{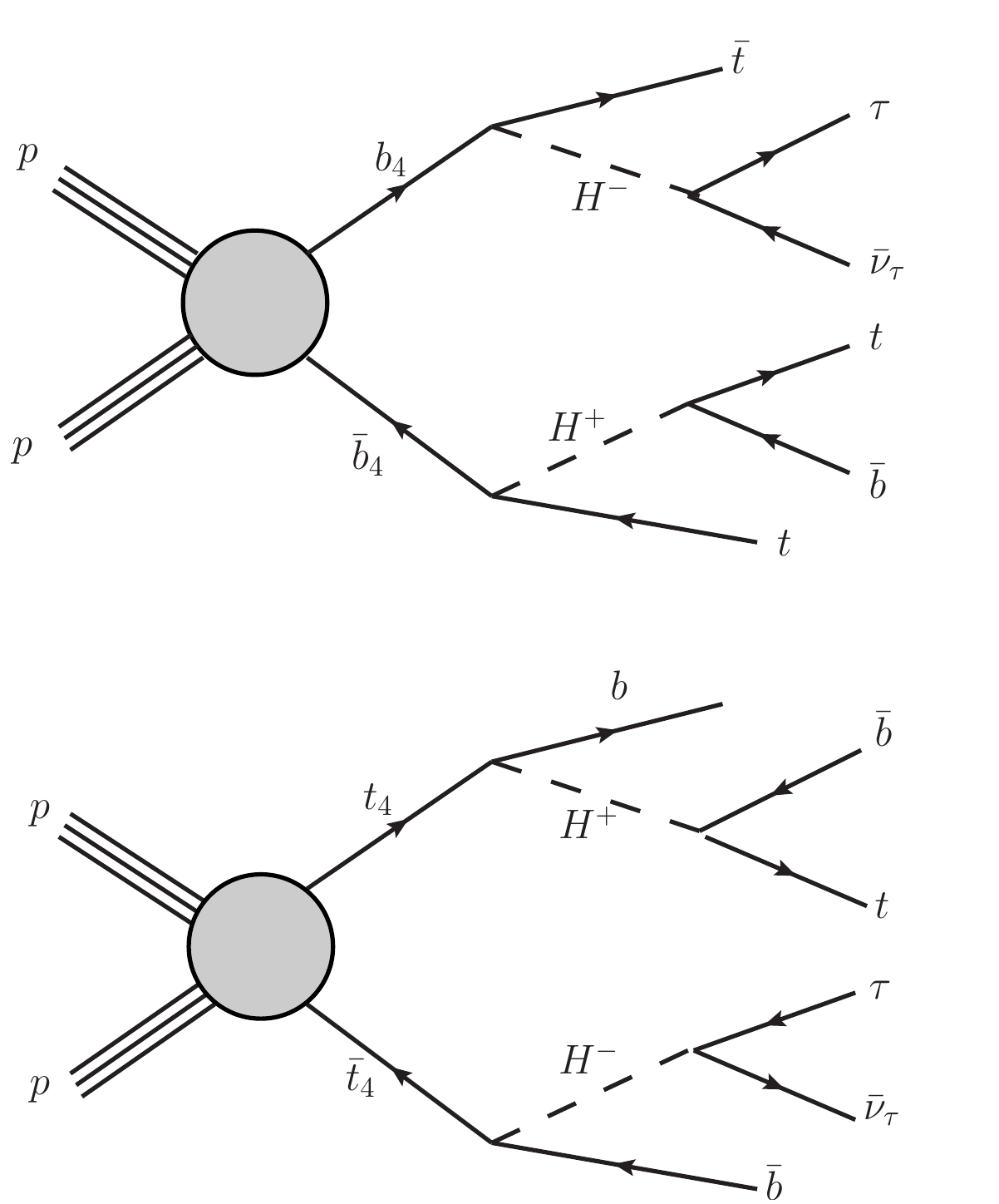}
\caption{Cascade decays of vectorlike quarks through heavy Higgs bosons leading to $4b+2\tau$ or $4b+1\tau$ final states. On the left and top right, we show possible decays of a bottom-like new quark through neutral and charged Higgses or both. On the bottom right we show the decay of a top-like new quark through charged Higgses.
}
\label{fig:diags}
\end{figure}

In this paper we present analyses based on final states where one of the heavy neutral or charged Higgs bosons decays to tau leptons, see figure~\ref{fig:diags}. Although in the type-II 2HDM the decay modes $H(A) \to \tau \tau$ and $H^\pm \to \tau \nu$ are subleading, they have the advantage of competing with significantly smaller background. Actually, in the type-II 2HDM,  $H(A) \to \tau \bar \tau$  leads to stronger limits~\cite{Aad:2020zxo, Sirunyan:2018zut} than  $H(A) \to b\bar b$~\cite{Sirunyan:2018taj, Aad:2019zwb}. We present search strategies for $4b+2\tau$ and $4b+1\tau$ signatures which are common to several possible decay chains involving both top and bottom quarks in final states and discuss the mass ranges of heavy Higgs bosons and new quarks that can be explored at the LHC as functions of branching ratios (BRs) in a model independent way. We find that the reach of the LHC for these decay modes is comparable to the reach for the similar processes with dominant decay modes of heavy Higgses, discussed in~\cite{Dermisek:2020gbr}. 
We further combine these analyses to study the ultimate reach of the high luminosity (HL) LHC and  interpret the sensitivity to charged and neutral Higgs bosons and vectorlike quarks in the type-II 2HDM. We find that the LHC reach for their masses extends to well above 2 TeV  in the case of an SU(2) doublet quark and to at least 1.8 TeV for a bottom-like SU(2) singlet quark in the whole range of $\tan\beta$ between 1 and 50.

The same or very similar signatures do not rely on type-II 2HDM and could be present in many models and frameworks with extended Higgs sectors and vectorlike quarks.  For scalars that couple to leptons more than in the type-II 2HDM the rates for the processes we consider could be further enhanced. These signatures are also relevant for models with $Z^\prime$ or $W^\prime$ with the neutral (charged) Higgs boson in figure~\ref{fig:diags} replaced by $Z^\prime$ ($W^\prime$). For example, a $Z^\prime$ with couplings to $b$ and a vectorlike quark was suggested in connection with the Z-pole anomalies~\cite{Dermisek:2011xu, Dermisek:2012qx} or tensions in rare $B$ decays~\cite{Kawamura:2019rth, Kawamura:2019hxp}.

Related signatures of cascade decays of heavy neutral Higgs bosons through vectorlike and SM quarks were recently studied in ref.~\cite{Dermisek:2019heo}. These decay modes, depending on the size and the structure of Yukawa couplings of vectorlike quarks, are relevant when vectorlike quarks are lighter than the new Higgs bosons. 

Possible embeddings into grand unified theories also suggest new leptons that form complete multiplets together with new quarks. Complementary signatures of heavy Higgses and vectorlike leptons were studied in refs.~\cite{Dermisek:2015oja, Dermisek:2015vra, Dermisek:2015hue, Dermisek:2016via, CidVidal:2018eel}. The SM with complete vectorlike families can provide an explanation of the observed
hierarchy of gauge couplings~\cite{Dermisek:2012as,Dermisek:2012ke}. In the MSSM with a complete vectorlike family, the seven largest couplings in the SM can be understood from the IR fixed point behavior provided that new quarks and leptons are in the multi-TeV range~\cite{Dermisek:2017ihj,Dermisek:2018hxq,Dermisek:2018ujw}, some of them possibly within the reach of the LHC. Vectorlike quarks around the same scales can also lead to more natural EW symmetry breaking~\cite{Dermisek:2016tzw,Cohen:2020ohi}, while vectorlike leptons can easily explain muon $g-2$ anomaly~\cite{Dermisek:2020cod, Dermisek:2021ajd, Dermisek:2014cia, Dermisek:2013gta, Czarnecki:2001pv, Kannike:2011ng}. There are many other observables related to vectorlike quarks and leptons in collider searches~\cite{Dermisek:2014qca,Kumar:2015tna,Bhattiprolu:2019vdu,Freitas:2020ttd,Bissmann:2020lge,Kawamura:2021ygg,Choudhury:2021nib}, $B$-meson anomalies~\cite{Raby:2017igl,Crivellin:2018qmi,Barman:2018jhz,Arnan:2019uhr,Kawamura:2019rth}, the Cabibbo angle anomaly~\cite{Endo:2020tkb,Crivellin:2020ebi}, and searches for dark matter partilces~\cite{Lu:2017uur,Kowalska:2017iqv, Calibbi:2018rzv,Jana:2020joi}.

This paper is organized as follows. In Sec.~\ref{sec:signal} we describe signal topologies and briefly review the model and expected pattern of branching ratios. In Section~\ref{sec:eventgen} we discuss search strategies for final states which contain two bottom quarks, and one or two tau leptons, and estimate the reach of the LHC. We summarize the main results in Sec.~\ref{sec:results}. Our conclusions are drawn in section~\ref{sec:conclusions}. Supplementary details of the analysis are provided in the appendix.

\section{Signal topologies and expected rates}
\label{sec:signal}
\begin{figure}[t]
\centering
\includegraphics[width=0.75\textwidth]{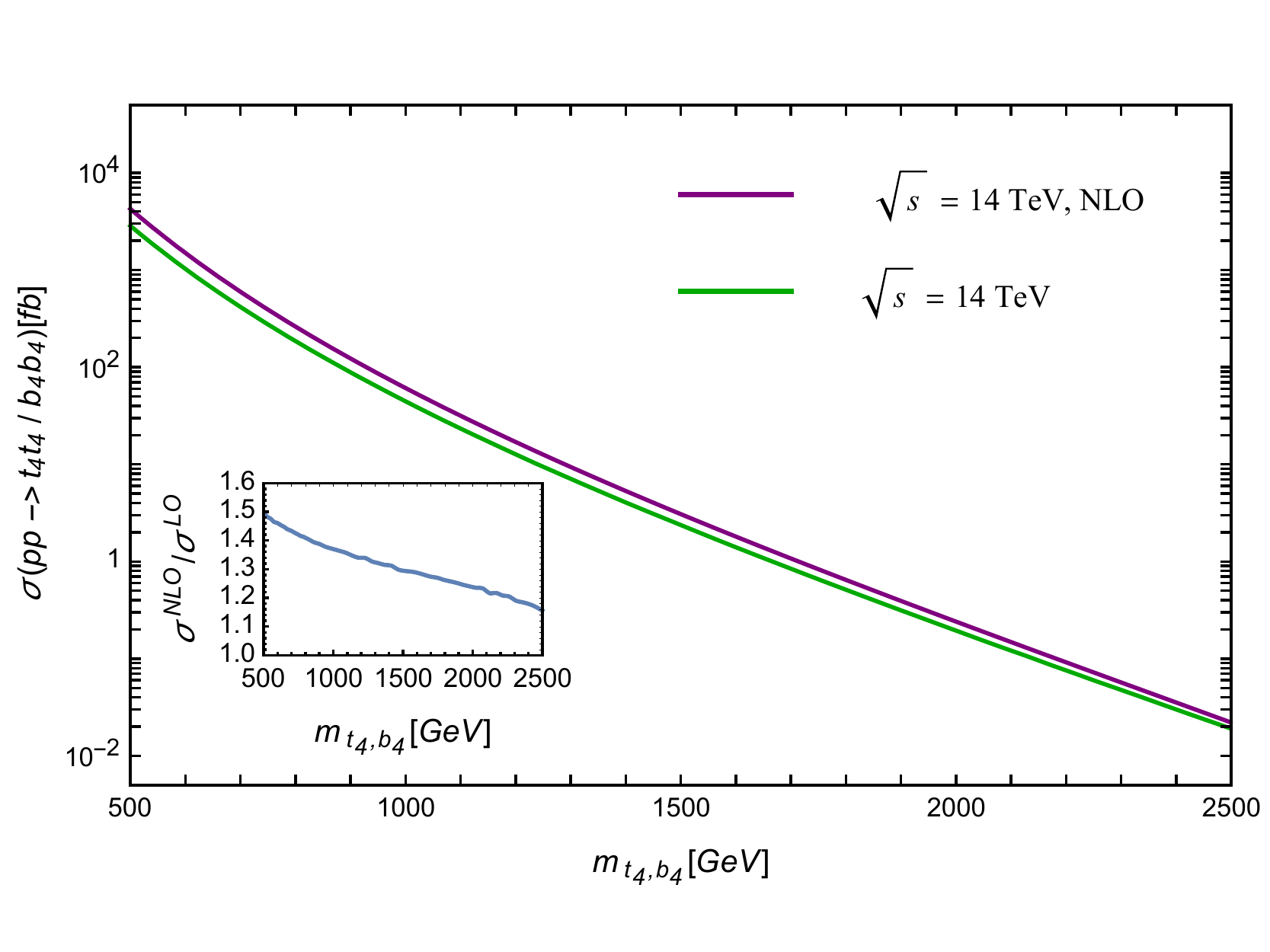}
\caption{NLO pair production cross section for top- or bottom-like vector-like quarks at the 14 TeV LHC with respect to the quark masses. In the inset we show the relative enhancement compared to the LO cross section. The LO cross sections have been obtained with {\tt MadGraph5}~\cite{Alwall:2014hca}, whereas the NLO cross sections were obtained with {\tt Top++}~\cite{Czakon:2011xx}.}
\label{fig:xsection}
\end{figure}

\begin{figure}[t]
\centering
\includegraphics[width=0.49\textwidth]{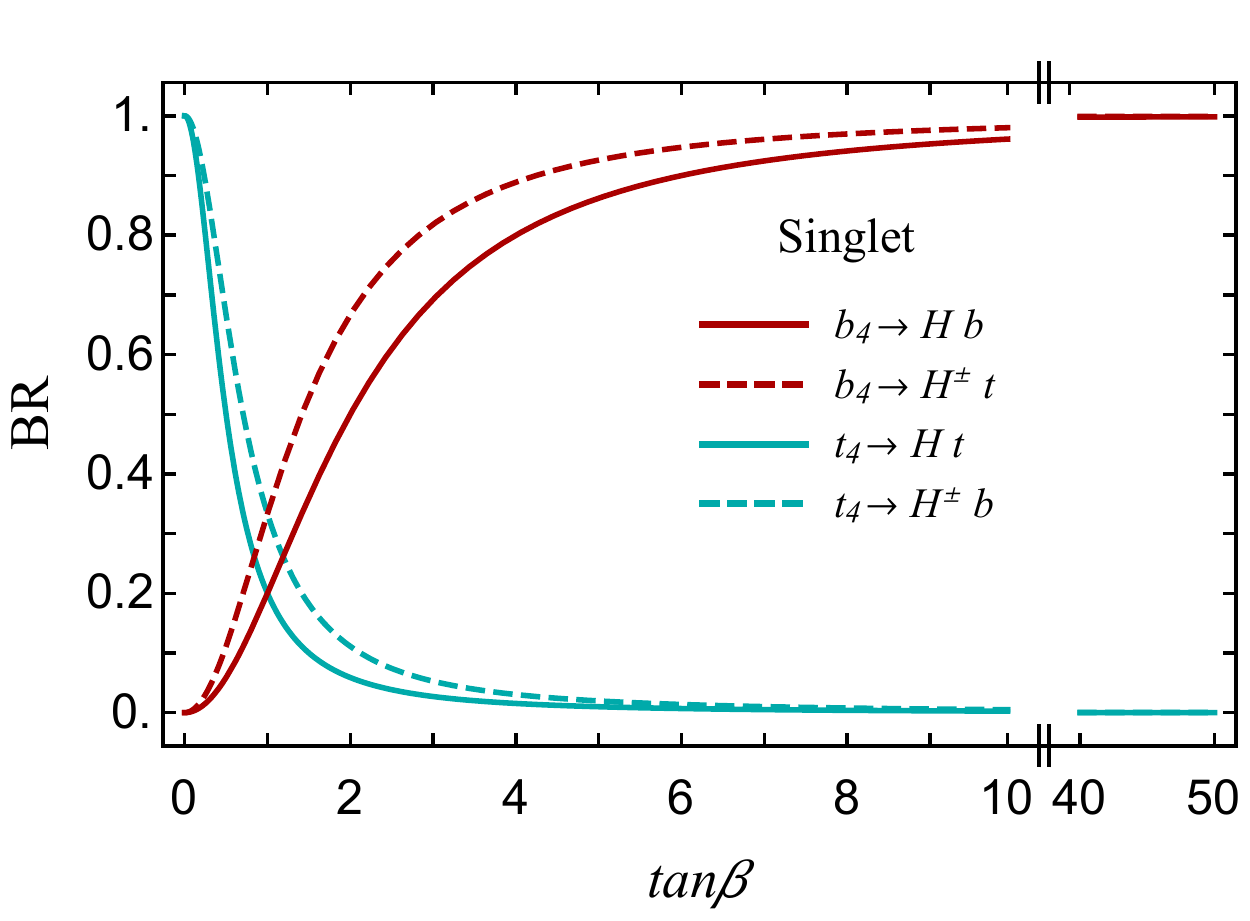}
\includegraphics[width=0.49\textwidth]{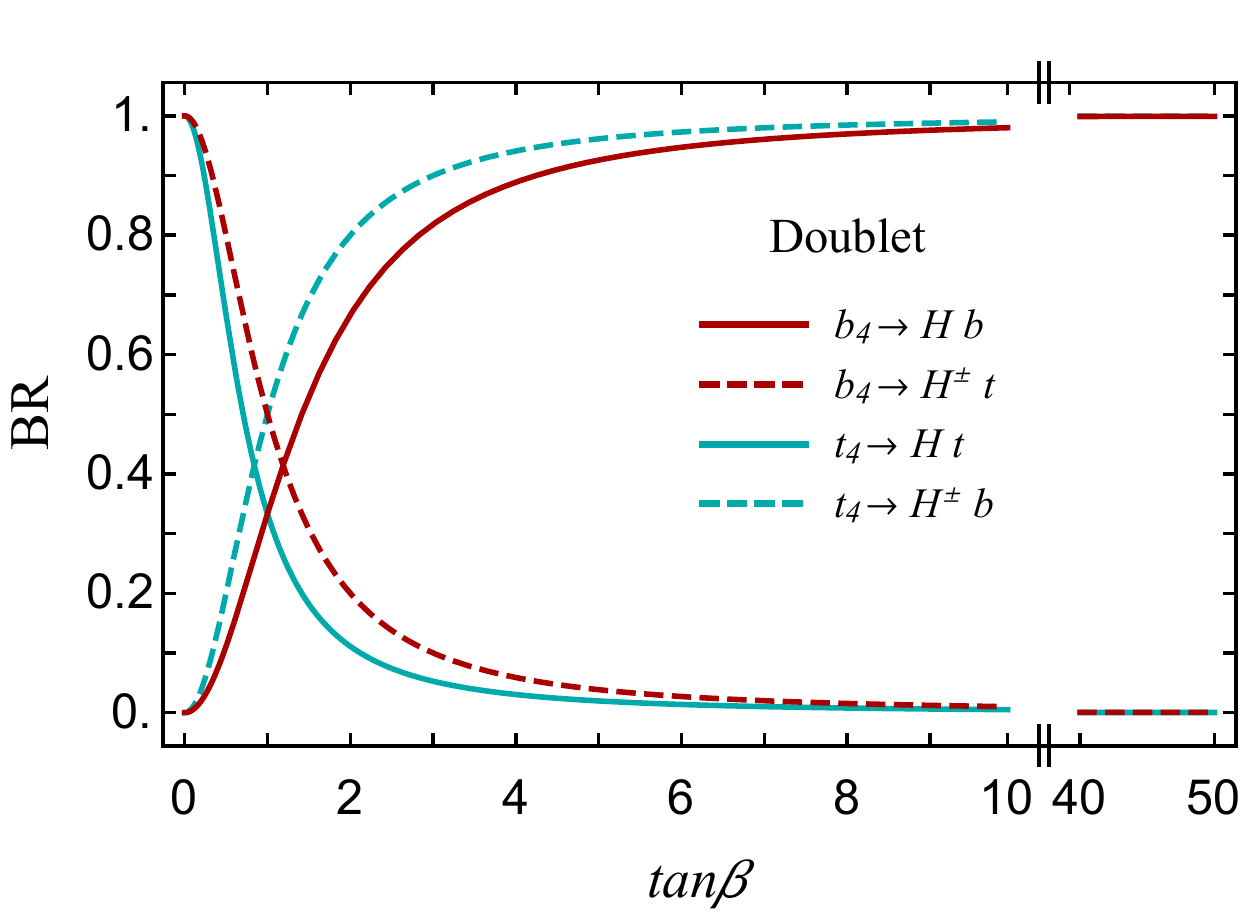}
\caption{Typical branching ratios of singlet-like (left) and doublet-like (right) $b_4$ and $t_4$ to charged and neutral heavy Higgs bosons with respect to $\tan\beta$ obtained from Table~6 of ref.~\cite{Dermisek:2019vkc}. The plots assume that a decay  through only one heavy Higgs boson is kinematically open and the remaining BRs are to $W$, $Z$, and $h$. The BRs to $A$ are the same as to $H$.}
\label{fig:singlet_doublet_BR}
\end{figure}

The $4b+2\tau$ and $4b+1\tau$ final states can be produced with large rates from cascade decays of new quarks through heavy Higgs bosons. The processes shown in figure~\ref{fig:diags} are expected to be the most sizable in the type-II 2HDM extended by $SU(2)$ doublet and singlet quarks  with the same quantum numbers as SM quarks and corresponding vectorlike partners. We label the lightest new mass eigenstates of bottom-like and top-like vectorlike quarks 
as $b_4$ and $t_4$. We assume that new quarks mix dominantly with the third generation of SM quarks, although the Yukawa couplings can be very small. We work in the alignment limit of the 2HDM so that the light CP-even Higgs boson, $h$, is SM-like. Details of the model have been presented in refs.~\cite{Dermisek:2019vkc, Dermisek:2019heo}. The pair production cross section of new quarks depends only on their mass, see figure~~\ref{fig:xsection}, while the typical  BRs of both new quarks and heavy Higgs bosons depend only on $\tan \beta$. 
Depending on the doublet/singlet nature of new quarks there are ranges of $\tan\beta$ where their decays  through heavy Higgses dominate the decays through $W$, $Z$, and $h$. Branching ratios of vectorlike quarks were studied in detail in ref.~\cite{Dermisek:2019vkc}.
The typical pattern of BRs, reproduced from table~6  of ref.~\cite{Dermisek:2019vkc}, is shown in figure~\ref{fig:singlet_doublet_BR} for singlet-like (left) and doublet-like (right) $b_4$ and $t_4$. The plots assume that a decay  through only one heavy Higgs boson is kinematically open and the remaining BRs are to $W$, $Z$, and $h$. The BRs to $A$ are the same as to $H$ (in this model $b_4$ has almost identical couplings to $H$ and $A$, see, for instance, appendix A.4 of ref.~\cite{Dermisek:2019vkc} and footnote~3 of ref.~\cite{Dermisek:2020gbr}). 
We see that a singlet-like $b_4$ can dominantly decay to both heavy neutral and charged Higgses at moderate to large $\tan\beta$,  while singlet-like $t_4$ decays to heavy Higgses can be sizable only for small $\tan\beta$. On the other hand, note that doublet-like
$b_{4}$ and $t_{4}$ can have large BRs to one of the neutral or charged Higgs bosons at any $\tan\beta$.

If decays through multiple heavy Higgs bosons are simultaneously kinematically open, the corresponding BRs are appropriately rescaled. 
Assuming the same masses of heavy Higgses, the typical ratios of partial widths of a singlet-like $b_4$ and $t_4$ to $H$, $A$, and $H^\pm$ are 1/4 : 1/4 : 1/2. Doublet-like $b_4$ and $t_4$  dominantly decay to the charged Higgs in a different region of $\tan\beta$ than to neutral Higgses. For more details, see the discussion in ref.~\cite{Dermisek:2020gbr}.
Subsequent decays of heavy Higgses follow the usual  pattern of type-II 2HDM.  Large BRs of heavy neutral Higgses to bottom quarks require medium to large $\tan\beta$ (see e.g. figure~3 of ref.~\cite{Dermisek:2019heo}). Moreover, branching ratios of both heavy neutral and charged Higgses into quarks and leptons are strongly correlated. For projected sensitivities at the HL-LHC we take ${\rm BR}(H / A \to b\bar b) = {\rm BR}(H^+ \to t\bar b) = 90\%$ and ${\rm BR}(H / A \to \tau^+ \tau^-) = {\rm BR}(H^+ \to \tau^+ \nu_\tau) = 10\%$ motivated by type-II 2HDM at large $\tan\beta$, however the sensitivities can be straightforwardly reinterpreted for any other pattern of BRs. 

The combination of $\tan \beta$ dependences of BRs of new quarks and heavy Higgses identifies processes shown in figure~\ref{fig:diags} as the most promising within the type-II 2HDM. However, the same or very similar signatures can arise in many other models with new quarks and scalars or new gauge bosons, $Z'$ or $W'$.
The search strategies discussed in following sections are thus not limited to the 2HDM.

\section{Event generation and selection}
\label{sec:eventgen}

All cascade decays in figure~\ref{fig:diags} yield final states with multiple $b$-jets and one or two $\tau$ leptons. Accordingly, we present two strategies in which we consider events with exactly one or two $\tau$-tagged jets. For each signal, events are generated using a {\tt FeynRules}~\cite{Degrande:2011ua} and corresponding {\tt UFO} model file based on the model described in detail in ref.~\cite{Dermisek:2019vkc}. Both signal and background events are generated using {\tt MadGraph5}~\cite{Alwall:2014hca} and are passed to  {\tt Pythia8}~\cite{Sjostrand:2006za, Sjostrand:2014zea} for showering and hadronization.  {\tt Delphes}~\cite{deFavereau:2013fsa} is used with the standard settings for the detector simulation; in particular for determining the origin of the $\tau$- and $b$-tagged jets. We note that the standard CMS simulation cards in {\tt Delphes} require $|\eta|<2.5$ in selecting jets that are either $b$- or $\tau$-tagged. However, the total efficiency for tagging $b$-jets is dependent on the $p_{T}$ (in units of GeV) of the observed jet, $\epsilon_{b}=\tanh(400/p_{T})85/(4+0.252 p_{T})$, while that of a $\tau$-jet is fixed with respect to $p_{T}$, $\epsilon_{\tau}=0.6$.

At the detector level, an advantage of search strategies based on $\tau$-jets is in avoiding large irreducible QCD backgrounds. As previously discussed~\cite{Goncalves:2015prv}, QCD processes involving final states with multiple $b$-jets can be easily faked by gluons splitting to a pair of $b$-quarks. $\tau$-jet final states avoid the need to include these large irreducible backgrounds allowing for simple cut-based strategies and confidence in the simulated background samples.  In addition, the analyses we present here rely largely on the ability to reconstruct decays of the top quark in order to suppress $t\bar t$ backgrounds which requires kinematic information of both $\tau$- and $b$-jets in the samples. Alternative strategies could be explored utilizing the leptonic decay modes of the $\tau$. However, these modes contribute only about $35\%$ of the total width and lack the features in the final state allowing to efficiently suppress the backgrounds originating from $t\bar t$.

The production cross sections $pp\to (b_4 \bar b_4, t_4 \bar t_4)$ depend only on the vectorlike quark mass and are presented in figure~\ref{fig:xsection} where we show the LO and NLO pair production cross section for the LHC running at 14 TeV. The NLO cross section has  been obtained using the public code {\tt Top++}~\cite{Czakon:2011xx} with standard settings. In the inset, we present the relative enhancement of the NLO cross section compared to the LO result. In the next two subsections we present the details of the $2\tau$ and $1\tau$ analyses.

\subsection{$2b+2\tau$ analysis}
\begin{figure}[t]
\centering
\includegraphics[width=0.49\textwidth]{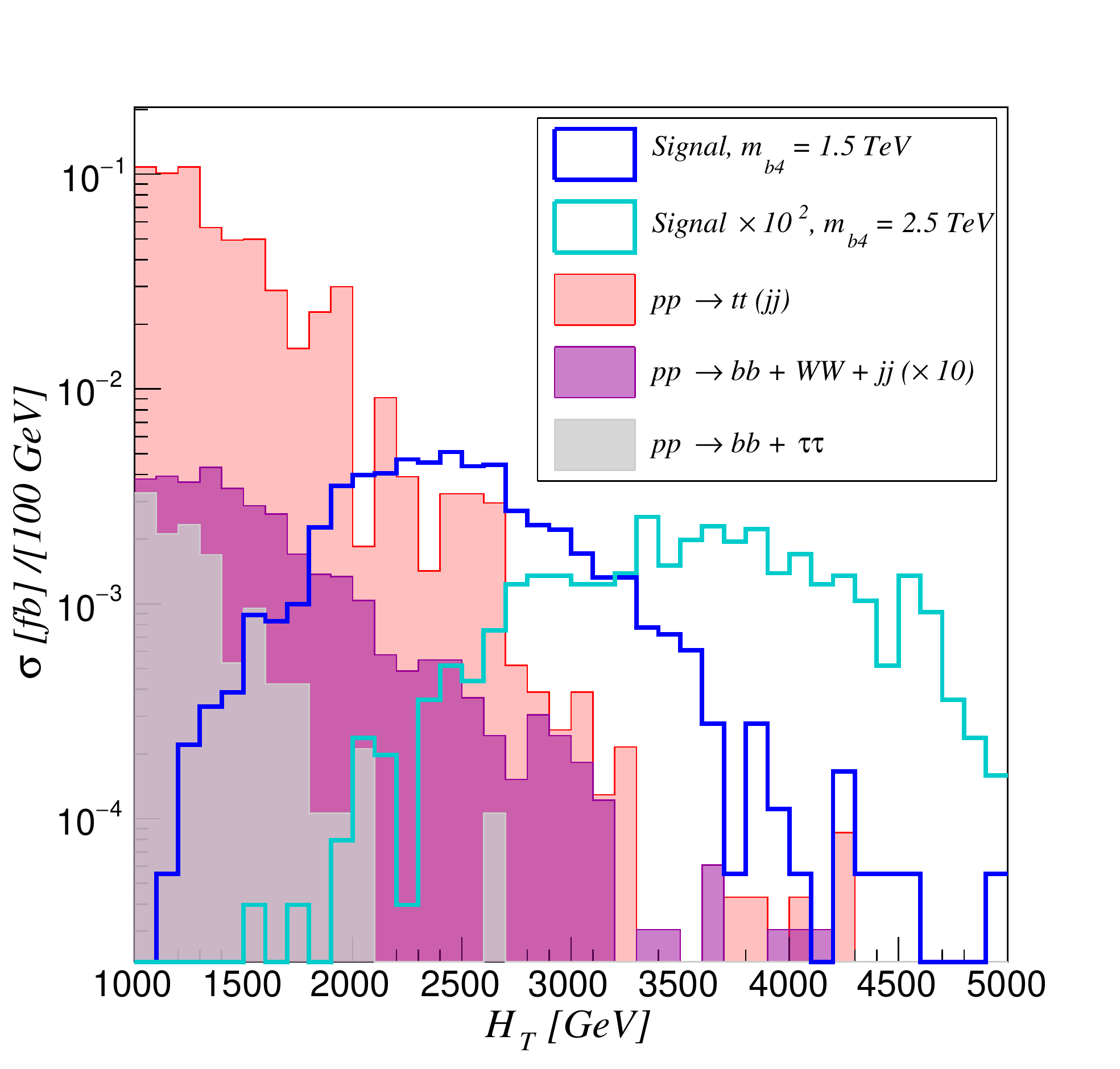}
\includegraphics[width=0.49\textwidth]{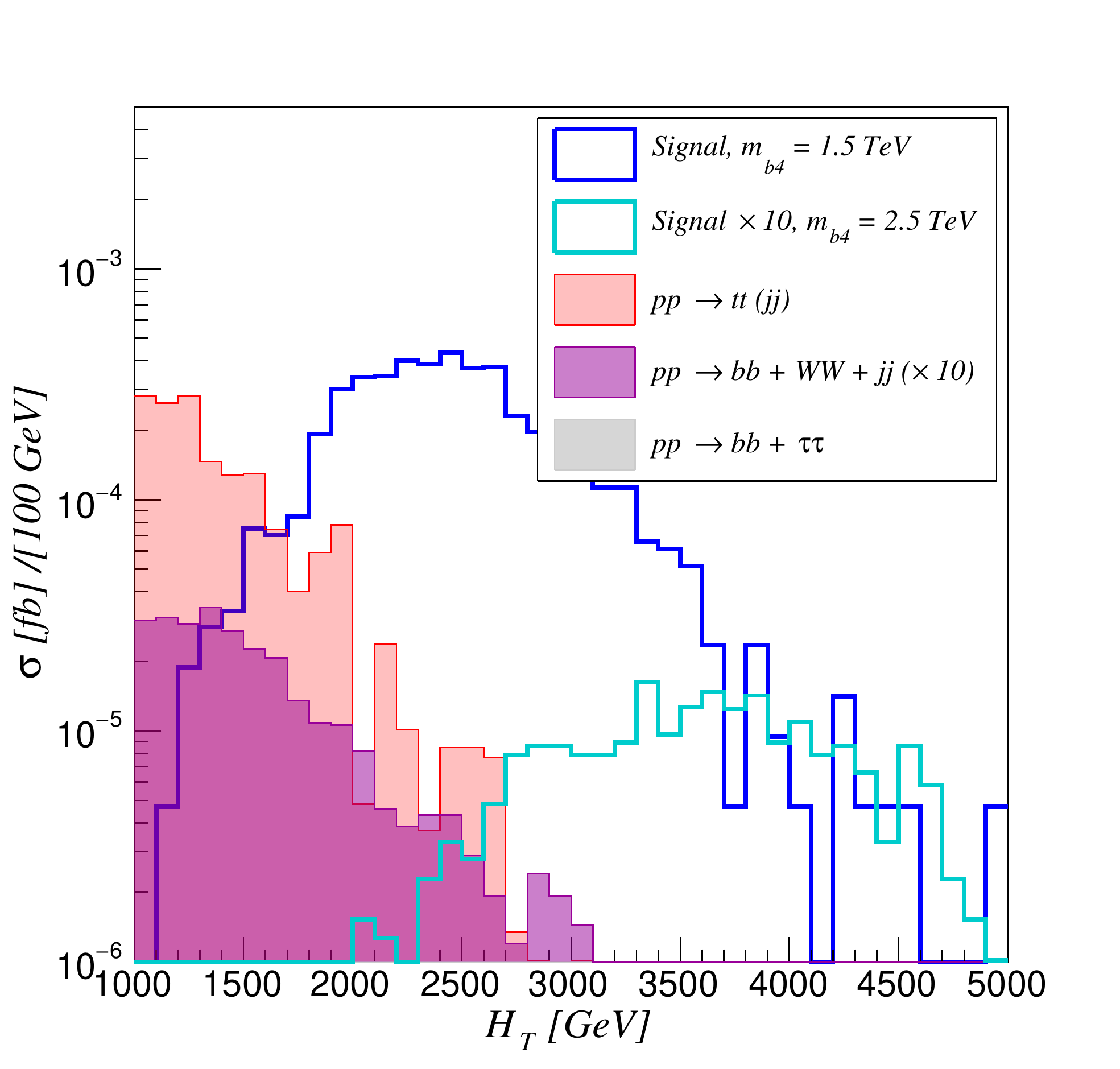}
\caption{Distribution of the total hadronic energy in the $2b+2\tau$ analysis before (left) and after (right) the anti-top cuts on $m_{\tau_{1}b_{i}}$ and $m_{\tau_{2}b_{i}}$. Signal distributions for $m_{b_{4}}=1.5$ and $2.5$ TeV are shown in blue and cyan respectively.
In order to make evident the position of the peak in the $H_{T}$ distributions we have rescaled the later by a factor of $10^{2}$ (left) and $10$ (right), respectively.
The red, green, magenta ($\times 10$), and gray regions show the background distributions from $tt+jj$, $tt$, $bb+WW+jj$, and $bb+\tau\tau$ respectively.}
\label{fig:6j_HT}
\end{figure}

For cascade decays leading to final states with two $\tau$ leptons, the initial event selection we require for this analysis is:
\begin{itemize}
\item Exactly two $\tau$-tagged jets with $p_T > 50$ GeV among the six highest $p_T$ jets;
\item At least four additional jets, any two of which are $b$-tagged, with $p_T > (200,100,100,50)$ GeV;
\item In order to suppress backgrounds from Standard Model Higgs and $Z$ boson, the invariant mass of the two $\tau$-tagged jets is required to be larger than 200 GeV. 
\end{itemize}
These initial cuts allow for two dominant sources of backgrounds: $b\bar b W^+ W^- (jj)$ and $b\bar b \tau^+ \tau^- (jj)$, where we include backgrounds with two additional jets denoted by $(jj)$, i.e. $b\bar b  \tau^+ \tau^- (jj)$ denotes the sum of  $b\bar b  \tau^+ \tau^-$ and $b\bar b  \tau^+ \tau^- +jj$ backgrounds.
We split the former in two parts corresponding to resonant $t\bar t (jj)$ where the $b$-$W^{\pm}$ pairs are produced from the decay of the top and the non-resonant remainder without tops in the decay chain. In figure~\ref{fig:6j_HT} (left), we show the transverse hadronic energy ($H_T$) distribution which we obtain after the initial selection described above; we show the relevant backgrounds and two signals corresponding to $m_{b_{4}}=1.5$ and 2.5 TeV (the signal cross sections correspond to maximal branching ratios and the latter has been rescaled by a factor $10^2$(left)). We see that even a moderate cut of $H_{T}>1500$ GeV alone can reduce most background events by roughly an order of magnitude compared to signal events even with lower masses. This kinematic advantage improves for heavier vectorlike quark masses. 

\begin{figure}[t]
\centering
\includegraphics[width=0.49\textwidth]{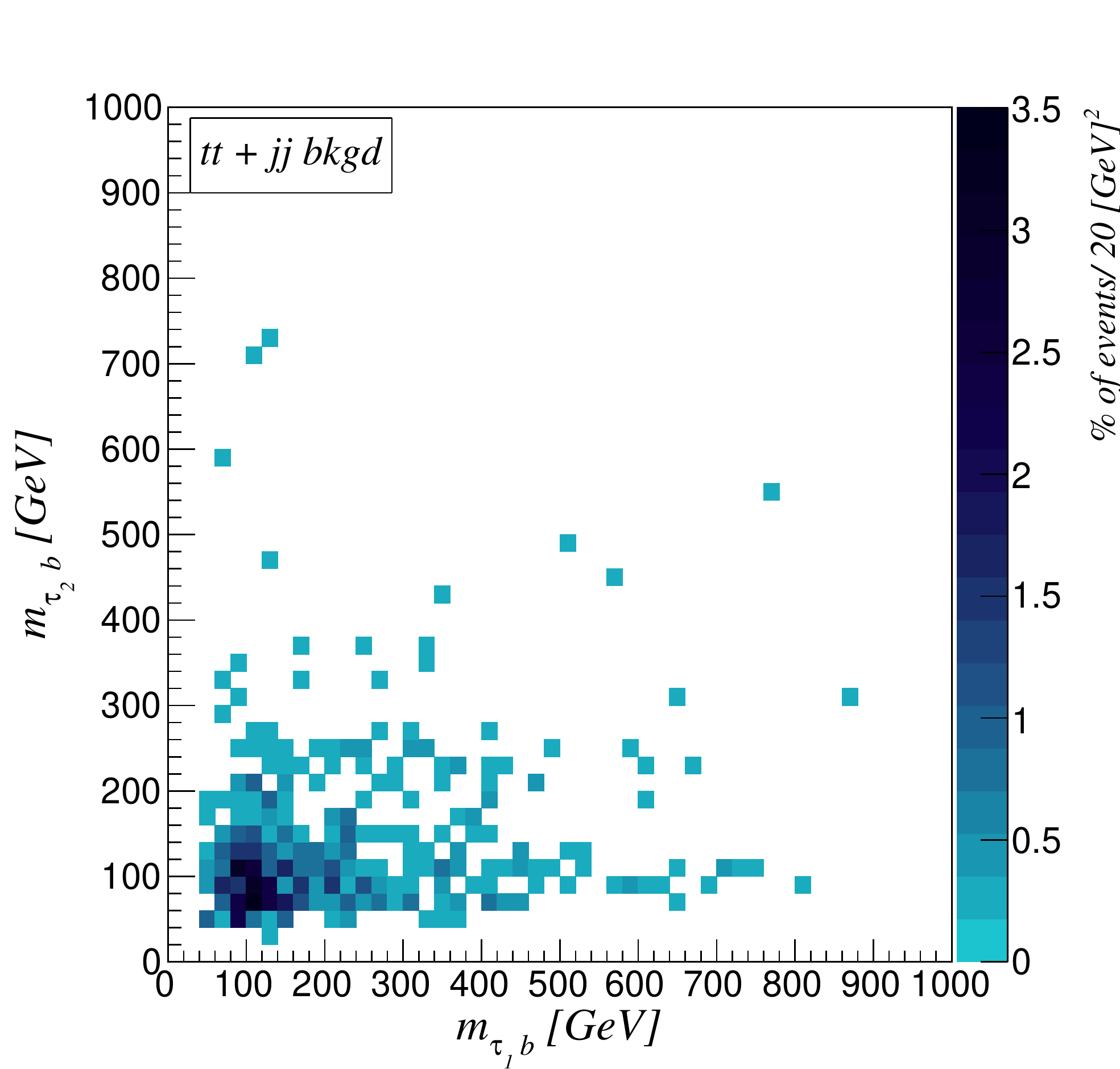}
\includegraphics[width=0.49\textwidth]{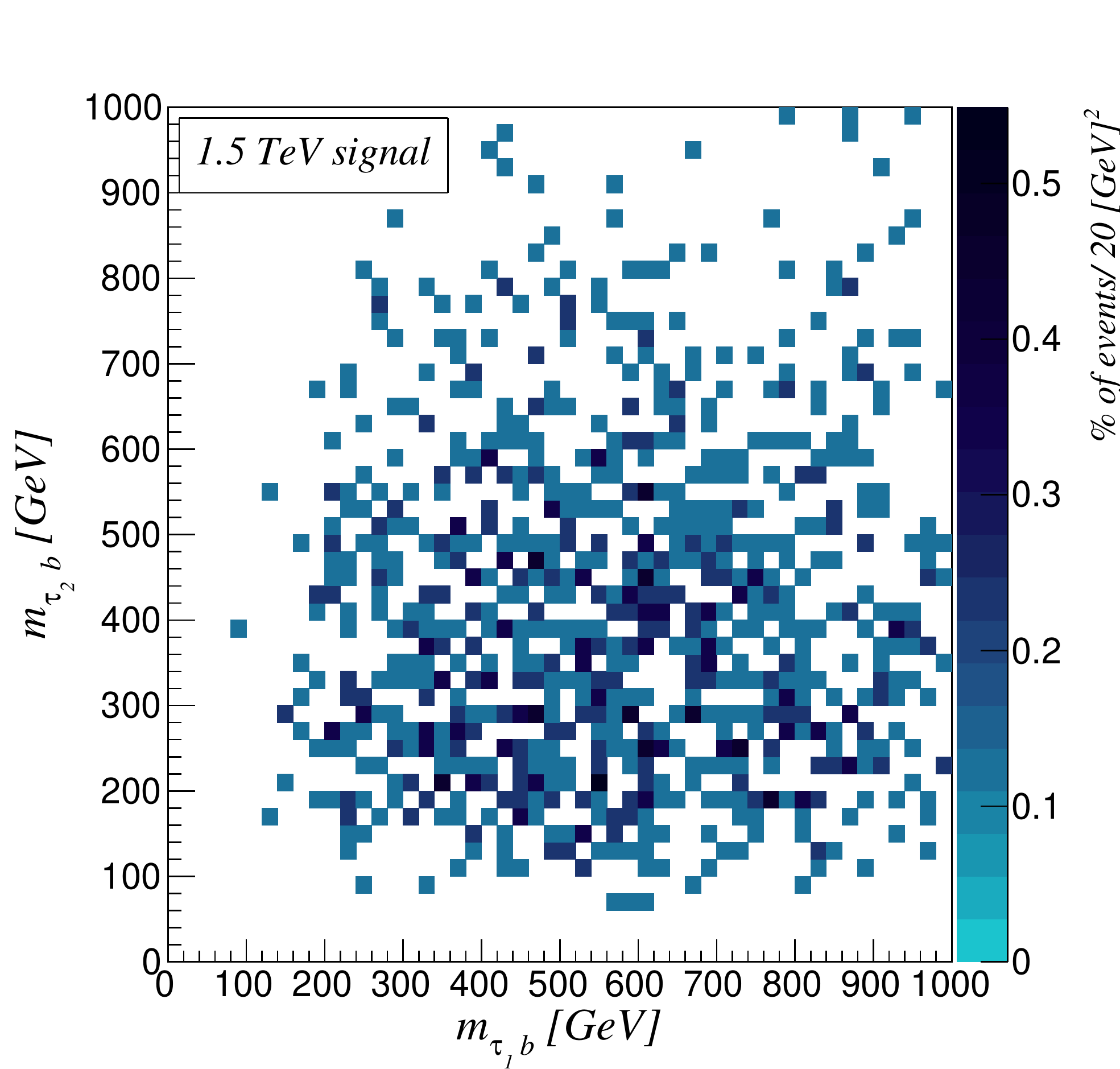}
\caption{{\bf Left:} The $tt+jj$ background distribution in the $m_{\tau_{1}b}$-$m_{\tau_{2}b}$ plane. {\bf Right:} Signal distribution in the $m_{\tau_{1}b}$-$m_{\tau_{2}b}$ plane assuming $m_{b_{4}}=1.5$ TeV.}
\label{fig:anti_top}
\end{figure}

It is clear that the most formidable background is $tt(jj)$. The top quark decays can be reconstructed by forming the minimum invariant mass between a $\tau$-tagged jet and a $b$-jet in the event. We begin by first forming the minimum of the invariant masses between the leading $\tau$-jet and any of the $b$-jets in the event $m_{\tau_{1}b_{i}}$ followed by the same for the next-to-leading $\tau$-jet and the remaining $b$-jets:
\begin{align}
m_{\tau_1 b} &=\underset{i}{\text{min}} \sqrt{(p(\tau_{1}) + p(b_{i}))^2} \; ,\\
m_{\tau_2 b} &=\underset{j \neq i_{\rm min}}{\text{min}} \sqrt{(p(\tau_{2}) + p(b_{j}))^2}\; ,
\label{eq:anti_top}
\end{align}
where $b_{i_{\min}}$ is the $b$-jet selected when constructing $m_{\tau_1 b}$. For $tt(jj)$ backgrounds, the distributions of these variables peak around the top quark mass, see figure~\ref{fig:anti_top} (left). By contrast, since the $\tau$-jets in signal events do not form a well-defined invariant mass with any $b$-jet in the event, the corresponding distribution shows no such correlation, see figure~\ref{fig:anti_top} (right) and this allows to efficiently distinguish the top decays. Thus, $tt(jj)$ can easily be cut away using an {\textit{anti-top}} cut requiring:
\begin{itemize}
\item $m_{\tau_1 b}$, $m_{\tau_2 b} > 200$ GeV.
\end{itemize}
The distributions we obtain after applying the anti-top cut are shown in figure~\ref{fig:6j_HT} (right), where we have rescaled the $m_{b_{4}}=2.5$ TeV signal instead by a factor of 10 to avoid overlap of the peak in the distribution with the legend. Note that while $\tau$-$b$ pairs in the $2b$+$2\tau$ background do not reconstruct the mass of the top quark, we find that the $m_{\tau_{1,2} b}$ invariant masses typically cluster even lower. Thus, events in this background in figure~\ref{fig:6j_HT} (right) are highly suppressed compared to the scale of the figure.
~\footnote{ Note that there are similar approaches using the information of top quark mass to subtract the $t \bar t$ background, e.g., in Ref~\cite{Kim:2019wns}. We believe adding these methods using the neural network would provide extra power in searching for our signals.}

Finally we apply a cut on $H_T$ which is optimized to the vectorlike fermion masses we consider:
\begin{itemize}
\item $H_T > [500,3000]$ GeV, depending on $m_{b_4,t_4}$.
\end{itemize}

\subsection{$2b+1\tau$ analysis}
\begin{figure}[t]
\centering
\includegraphics[width=0.49\textwidth]{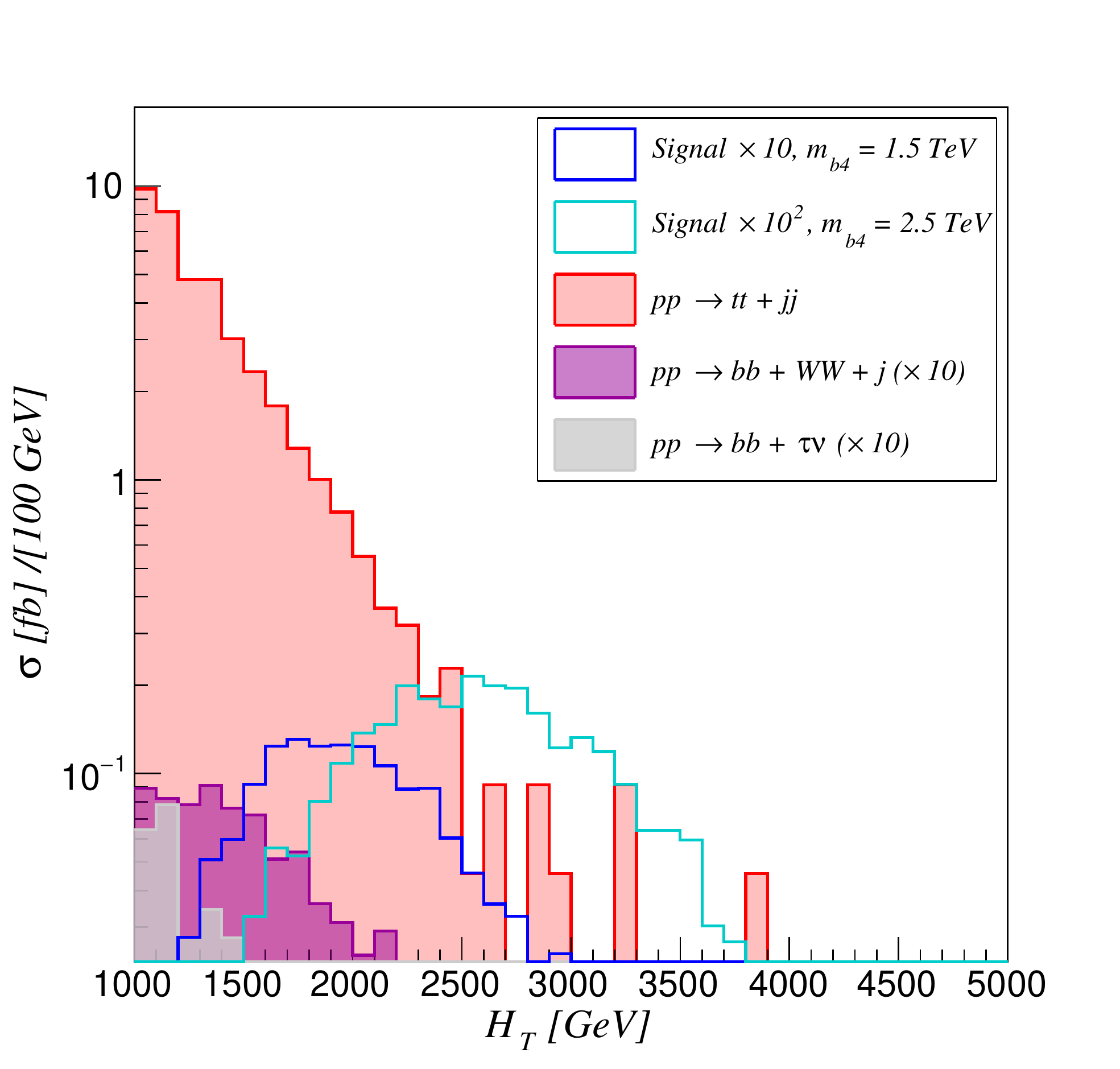}
\includegraphics[width=0.49\textwidth]{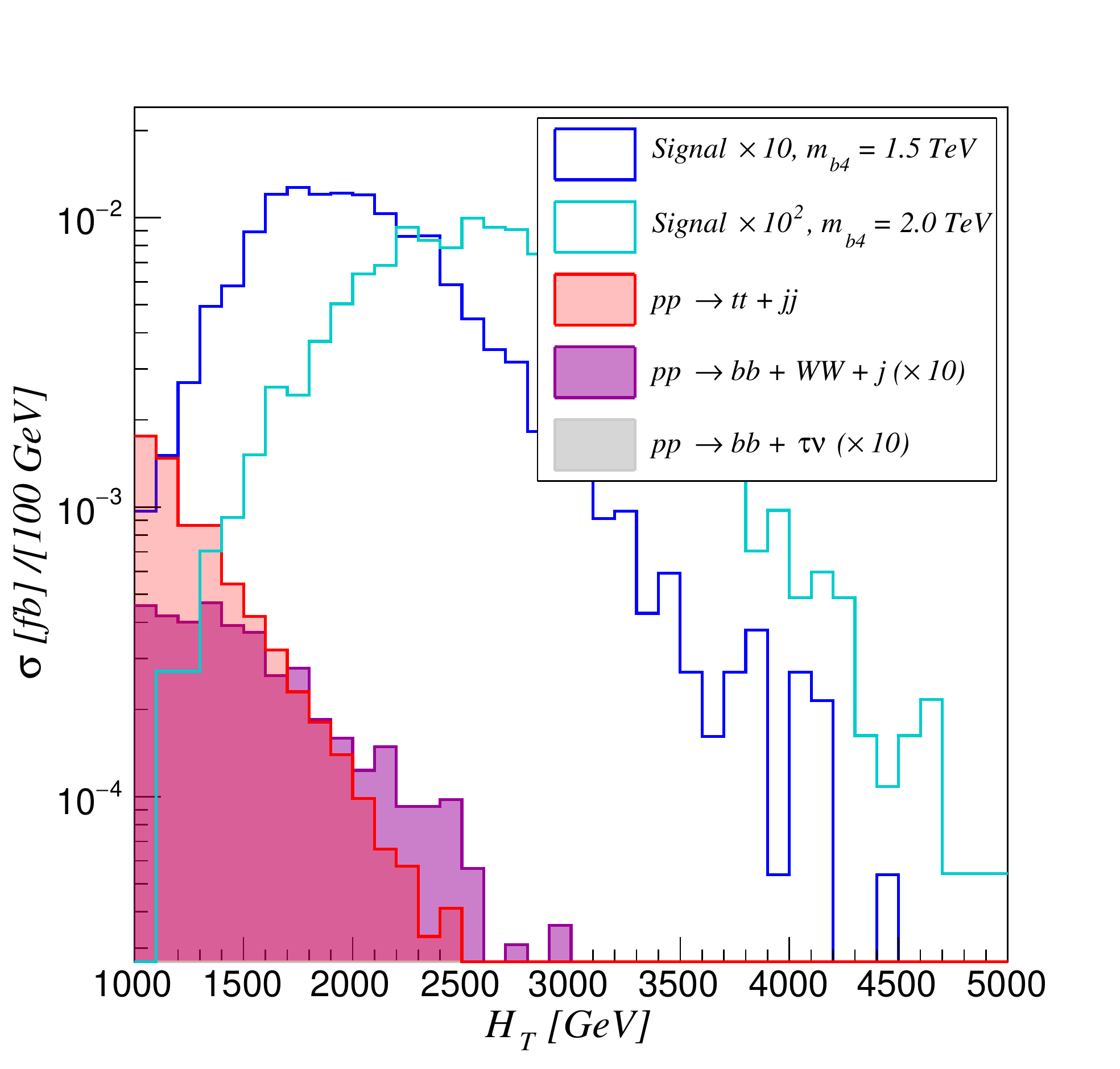}
\caption{Distribution of $H_{T}$ in the $2b+1\tau$ analysis before (left) and after (right) the anti-top cut on $m_{\tau b_{i}}$. Signal distributions for $m_{b_{4}} = 1.5$ and 2.0 TeV are shown in blue and cyan respectively, where the latter has been scaled by a factor of $10^{2}$. Background distributions for $tt(jj)$, $bb+WW+j (\times 10)$, and $bb+\tau\nu(\times 10)$ are shown with corresponding colors as in figure~\ref{fig:6j_HT}.}
\label{fig:5j_HT}
\end{figure}

For cascade decays leading to final states with a $\tau$ and $\nu_{\tau}$ pair, the initial event selection we require for this analysis is:
\begin{itemize}
\item Exactly one $\tau$-tagged jet with $p_T > 50$ GeV among the five highest $p_T$ jets;
\item At least four additional jets, any two of which are $b$-tagged, with $p_T > (200,100,100,50)$ GeV;
\item A lower bound on the total missing transverse energy $\slashed{E}_{T}>350$ GeV to suppress leptonic decays of the $W$ boson in the backgrounds;
\item In order to suppress backgrounds from tops, the minimum invariant mass, $m_{\tau b}$, between the $\tau$-tagged jet and any of the two $b$-tagged jets is required to be larger than 300 GeV.
\end{itemize}
Since we have only one $\tau$-tagged jet we remove part of the $t\bar t(jj)$ background by strengthening the cut on $m_{\tau b}$. These initial cuts allow for backgrounds from resonant $t\bar t (jj)$, non-resonant $b\bar b W^+ W^- (j)$ and $b\bar b \tau \nu_\tau$. In the two panels of figure~\ref{fig:5j_HT} we present the $H_T$ distribution of signal and backgrounds before (left) and after (right) the anti-top cut $m_{\tau b} > 300$ GeV. The comparison highlights the impact of this cut on the resonant $t\bar t (jj)$ background. The suppression of the $b\bar b \tau \nu_\tau$ background is due to a similar effect as for  $2b$+$2\tau$ in the previous analysis.

\begin{figure}[t]
\centering
\includegraphics[width=0.49\textwidth]{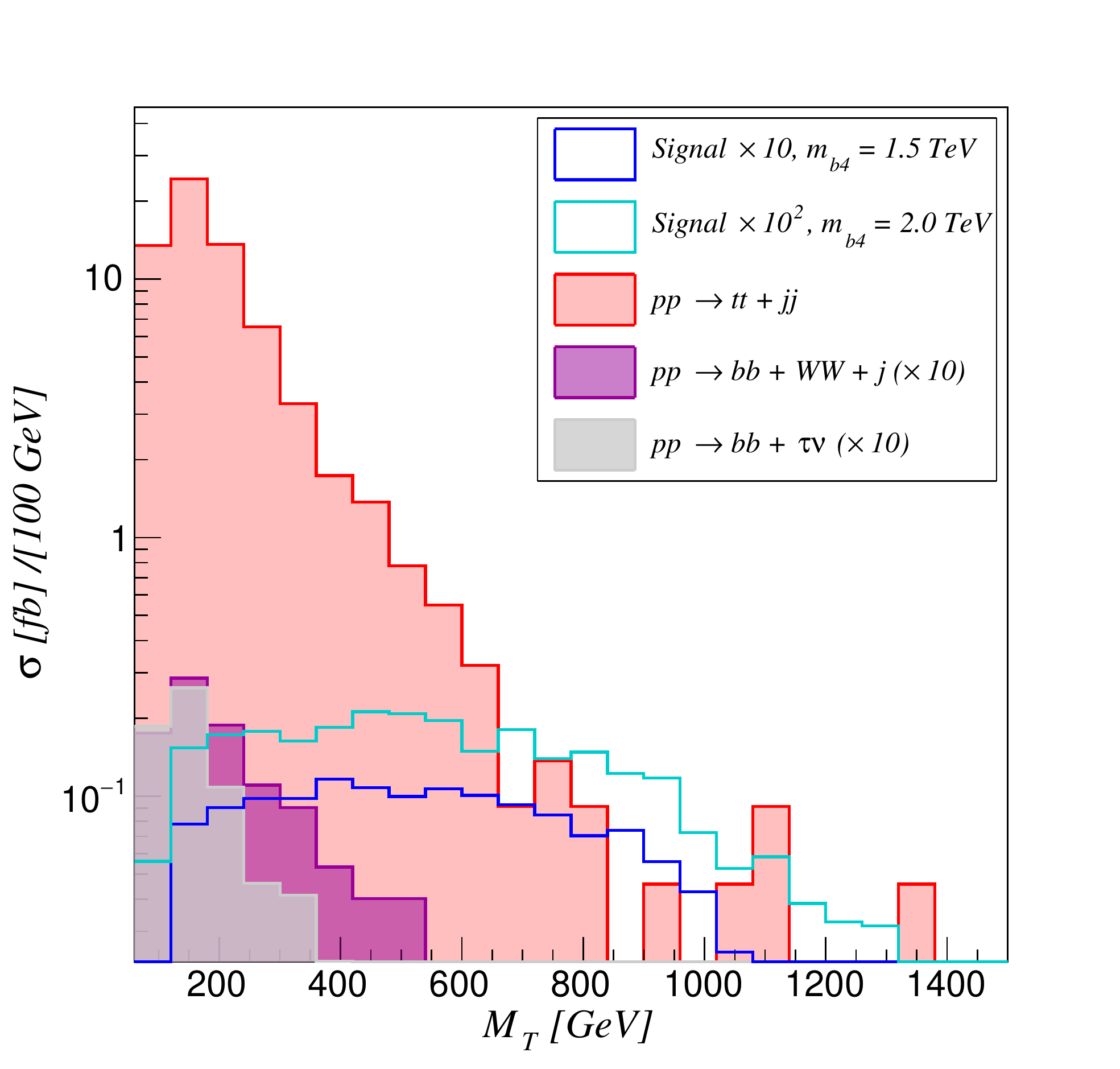}
\includegraphics[width=0.49\textwidth]{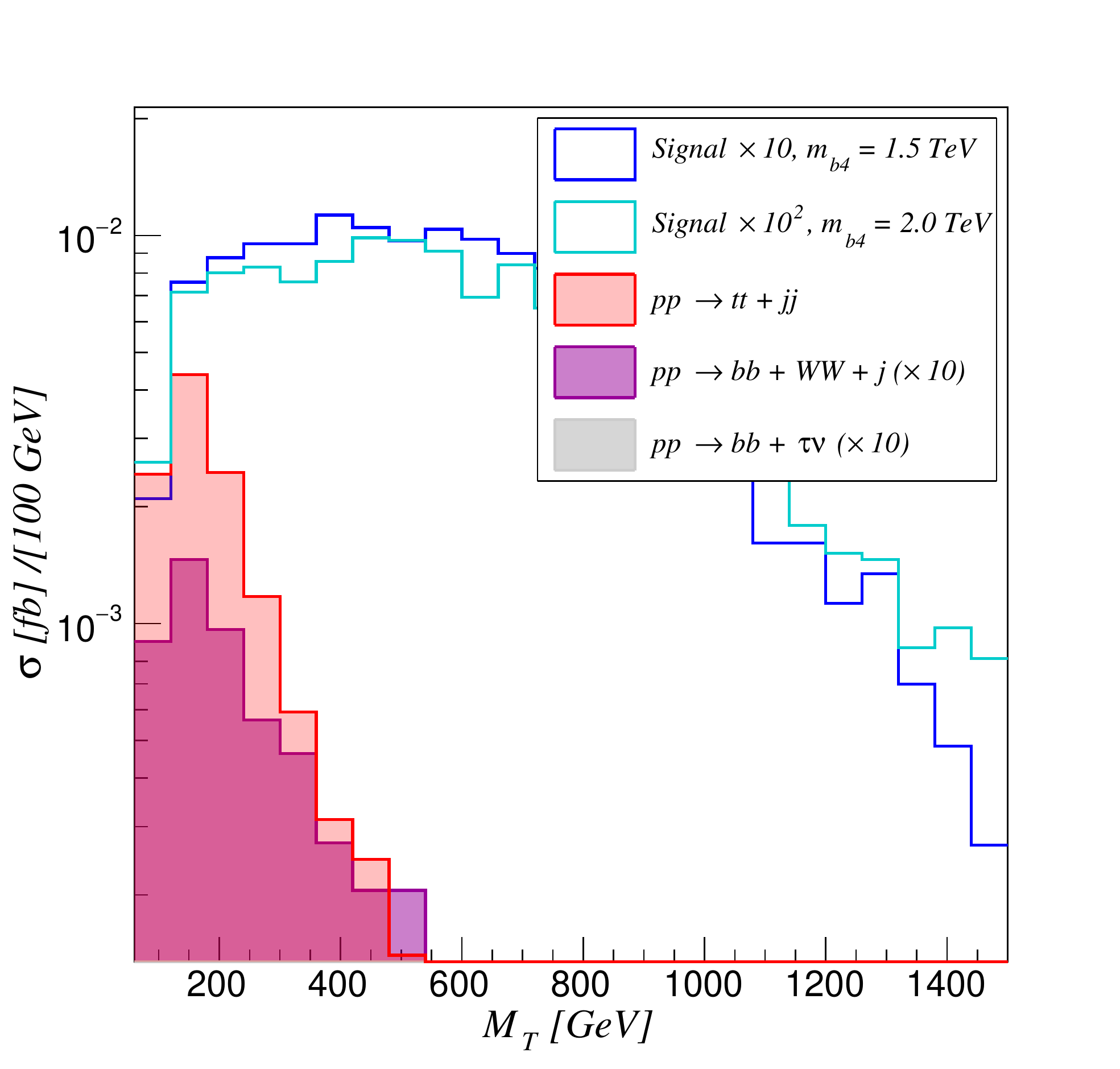}
\caption{Distribution of the transverse mass $M_T$ in the $2b+1\tau$ analysis before (left) and after (right) the anit-top cut on $m_{\tau b_{i}}$. Color coding is the same as in Fig.~\ref{fig:5j_HT}.}
\label{fig:MT}
\end{figure}

The $2b+1\tau$ analysis is aimed at constraining signal events in which the $\tau$ originates from the charged Higgs decay $H^\pm \to \tau^\pm \nu_\tau$, implying the presence of significant missing energy. Apart from a cut on $\slashed{E}_{T}$, it is expected that the correlation between the $p_{T}$ of the $\tau$-tagged jet with the $\slashed{E}_{T}$ in the events will differ in the signal compared to the background due to the heavy intermediate particle. We take advantage of this feature of the signal by considering the {\textit{transverse mass}} of the $\tau - \slashed{E}_{T}$ system:
\begin{align}
M_{T}=\sqrt{\left(|\vec p_{T}(\tau)| + |\vec{\slashed{p}}_{T}|\right)^{2} - \left|\vec{p}_{T}(\tau) + \vec{\slashed{p}}_{T}\right|^{2}}.
\end{align}
In figure~\ref{fig:MT}, we show the distribution of $M_{T}$ in the $2b+1\tau$ analysis for signal events from $b_{4}$ decays and relevant backgrounds. While the distribution for signal is somewhat spread, hindering a strong cut, large portions of the background can be reduced by a moderate cut on $M_{T}$. 

The final cuts we apply are on $H_T$ and $M_T$ which are optimized to the vectorlike fermion mass we consider:
\begin{itemize}
\item $M_{T} > [100,700]$ GeV, depending on $m_{b_4,t_4}$;
\item $H_{T} > [500,3000]$ GeV, depending on $m_{b_4,t_4}$.
\end{itemize}

\section{Results}
\label{sec:results}

\subsection{Model-independent sensitivities of $2b+2\tau$ and $2b+1\tau$ analyses }

\begin{figure}[t]
\centering
\includegraphics[scale=1]{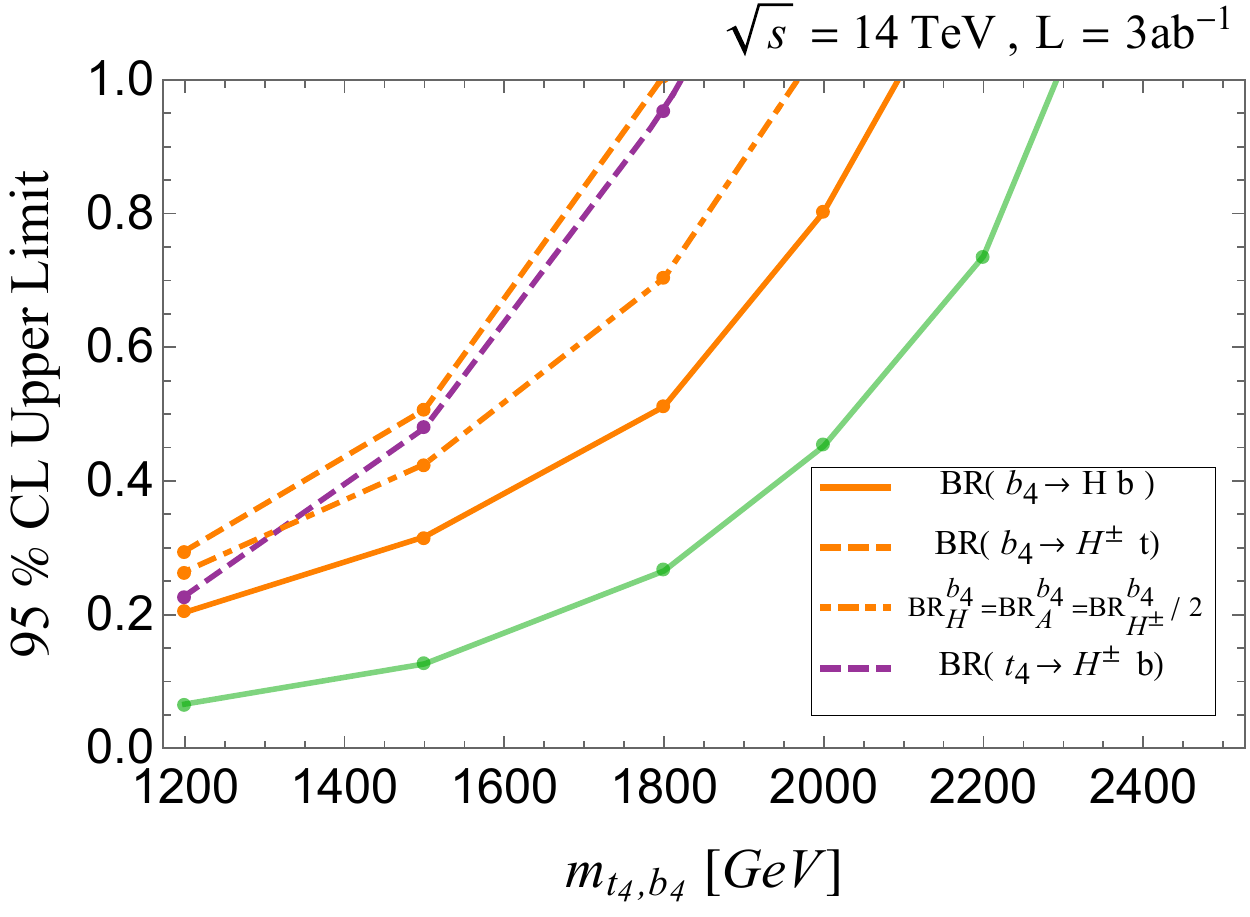}
\caption{95\% CL {projected upper limits on $t_{4}$ and $b_{4}$ cascade decays leading to $2b+2\tau$ or $2b+1\tau$ final states for the 14 TeV HL-LHC. For comparison, in green we show the projected limits} on ${\rm BR}(b_{4}\to Hb)$ using the 5b analysis in ref.~\cite{Dermisek:2020gbr} updated with the NLO production cross section and assuming ${\rm BR}(H\to b\bar{b})=0.9$.}
\label{fig:BR_bounds}
\end{figure}

We begin by assuming that only one heavy Higgs channel is kinematically open and proceed to find sensitivities of the HL-LHC to  $b_{4}\rightarrow H b$, $b_{4}\rightarrow H^\pm t$, and $t_{4}\rightarrow H^\pm b$. The $b_{4}\rightarrow H b$ can be constrained by the $2b+2\tau$ analysis and the HL-LHC sensitivity is given by the solid orange curve in figure~\ref{fig:BR_bounds}. Constraining branching ratios into the charged Higgs require the weaker $2b+1\tau$ analysis which yields the dashed orange ($b_4$) and purple ($t_4$) projected limits in figure~\ref{fig:BR_bounds}. These expected bounds have been calculated for $M_{H,H^\pm} = 1$ TeV. However, we have verified that allowing for heavier Higgs bosons, up to 200 GeV lighter than the decaying vectorlike quark, does not have a significant impact on the bounds. We see that sensitivities up to 2.1 TeV and 1.8 TeV are possible for decays into neutral heavy Higgs (from $b_4$) and charged Higgs (both $b_4$ and $t_4$). The absence of a signal for lower quark masses can be interpreted as an upper limit on the branching ratio of a new quark to heavy Higgses.

Note that the plotted sensitivities implicitly assume ${\rm BR}(H / A \to b\bar b) = {\rm BR}(H^+ \to t\bar b) = 90\%$ and ${\rm BR}(H / A \to \tau^+ \tau^-) = {\rm BR}(H^+ \to \tau^+ \nu_\tau) = 10\%$ motivated by type-II 2HDM at large $\tan \beta$.  In models with different patterns of branching ratios the plotted sensitivities are on ${\rm BR}(b_4 \to H b) \times \sqrt{ [{\rm BR} (H\to b\bar b)/0.9] \times[ {\rm BR} (H/A \to \tau^+\tau^-)/0.1] }$ and with a similar modification for the decays through the charged Higgs. 

In the case that $b_{4}$ is an $SU(2)$ singlet, the rates for decays through neutral Higgses, charged Higgses or mixed decays through both charged and neutral Higgses (see figure~\ref{fig:diags} bottom left) can  be simultaneously large. 
In this scenario the $2b+2\tau$ analysis is sensitive to the top-left and bottom-left diagrams in figure~\ref{fig:diags}, while the $2b+1\tau$ to the top-right and bottom diagrams. To set a limit on the overall
 decay of $b_{4}$ to heavy Higgses, we assume that $H$, $A$ and  $H^\pm$  are degenerate in mass and that the BRs of $b_4$ into them are typical, 1/4 : 1/4 : 1/2,  as discussed in section~\ref{sec:signal}.
The HL-LHC sensitivity to the combined branching ratio into heavy Higgses resulting from a statistical combination of the $2b+2\tau$ and $2b+1\tau$ analyses is given by the dot-dashed orange curve. The procedure we adopt for the statistical combination of the two analyses is based on Poisson statistics and is summarized in appendix~\ref{app:poisson}. Since the $2b+2\tau$ analysis has superior reach than the $2b+1\tau$ one, these bounds are in between those we obtained from analyses in which decays into only one Higgs are kinematically allowed. We see that we achieve sensitivities close to 2 TeV.

For comparison, in green we also show the projected limits on ${\rm BR}(b_{4}\to Hb)$ using the 5b analysis in ref.~\cite{Dermisek:2020gbr} updated with the NLO production cross section and assuming ${\rm BR}(H\to b\bar{b})=0.9$. We see that despite an order of magnitude smaller branching ratio, the $H \to \tau^+ \tau^-$ leads to a similar reach in masses of new quarks and heavy Higgses at the HL-LHC. Furthermore, in case that the signal is observed in either of the final state, searching for the complementary signal in other decay modes would be crucial to uncover the nature of new physics.

\subsection{Reach of HL-LHC for new quarks and heavy Higgses in type-II 2HDM}
\begin{figure}[t]
\centering
\includegraphics[scale=0.5]{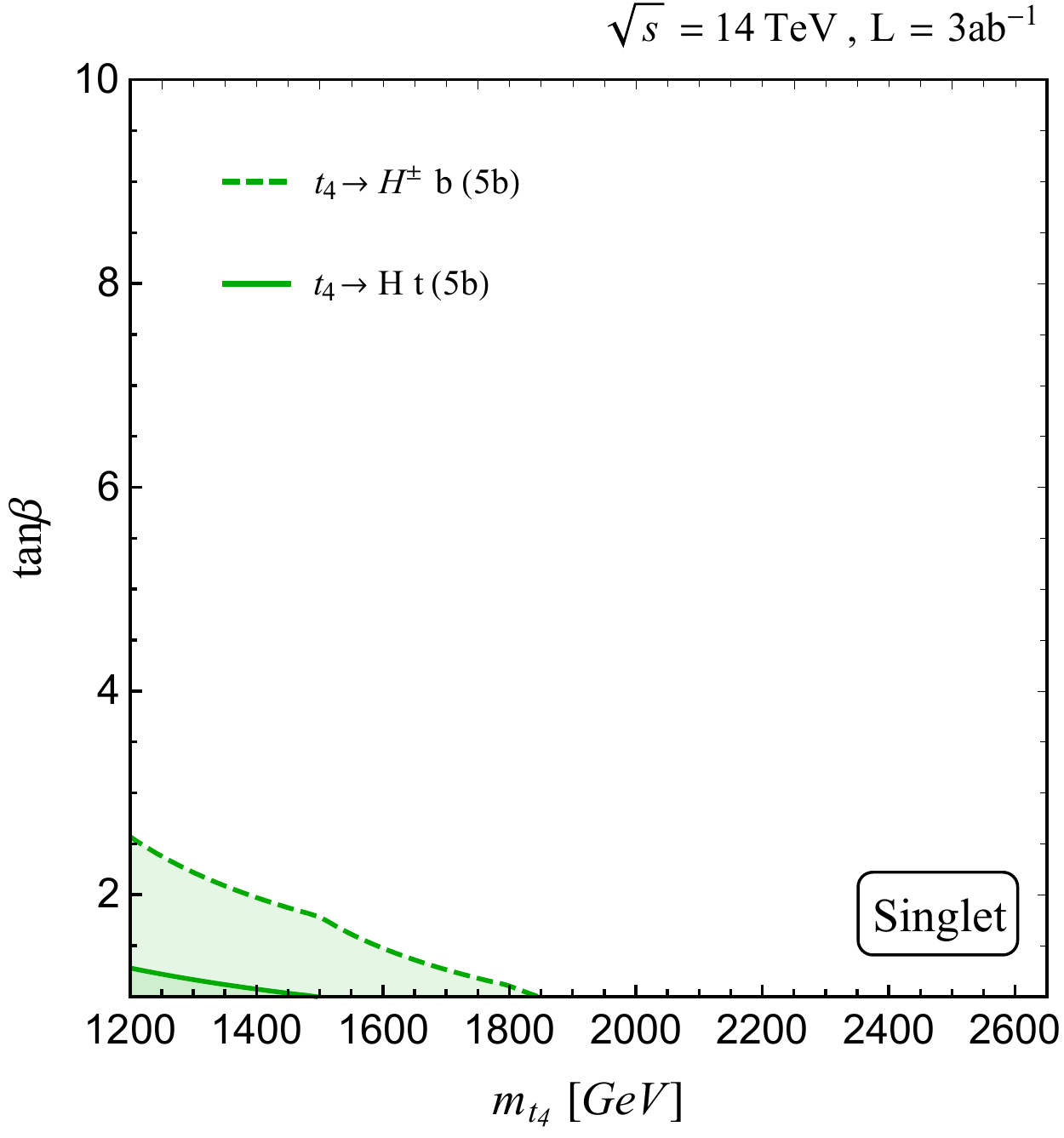}
\includegraphics[scale=0.5]{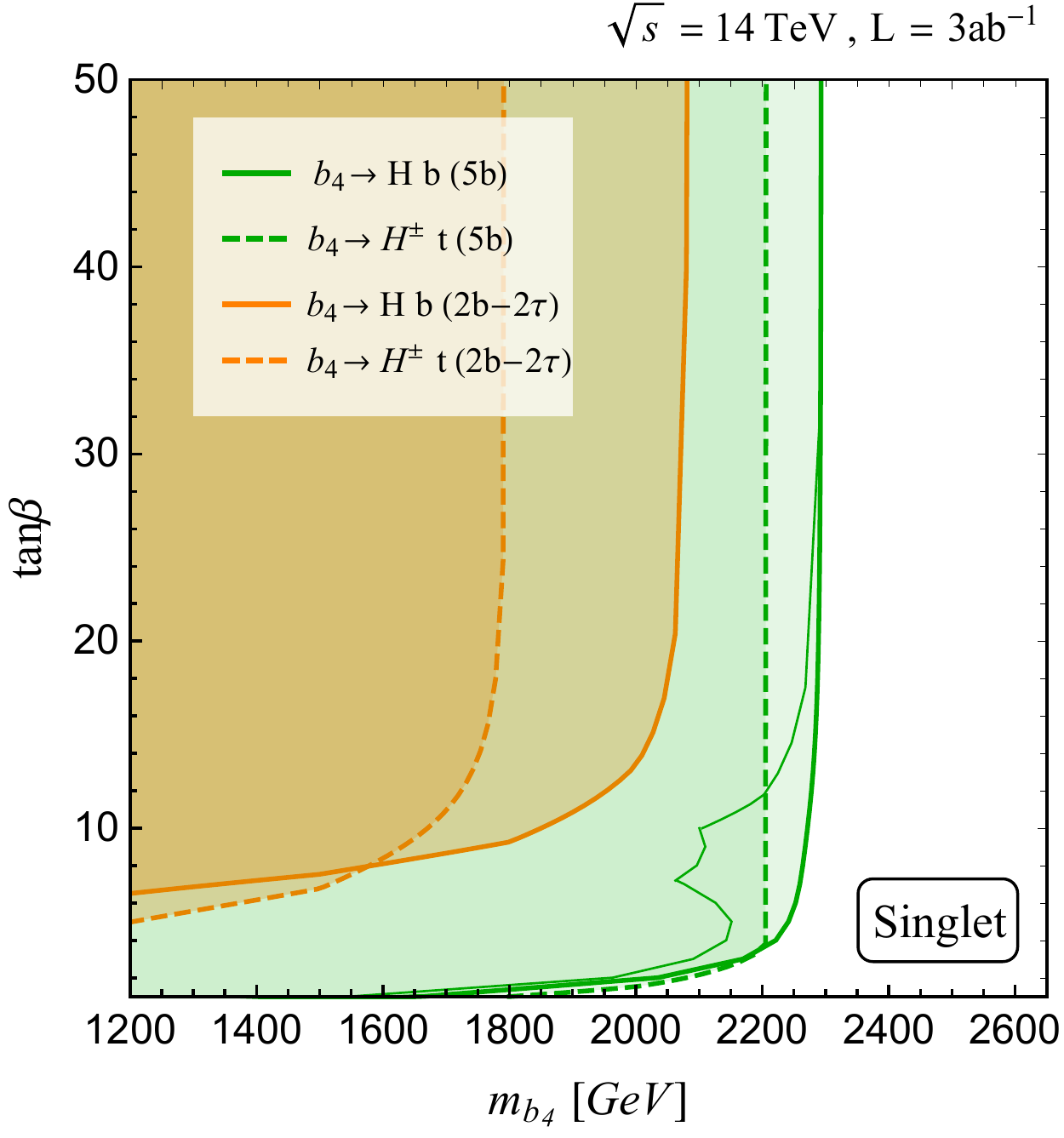}\\
\includegraphics[scale=0.5]{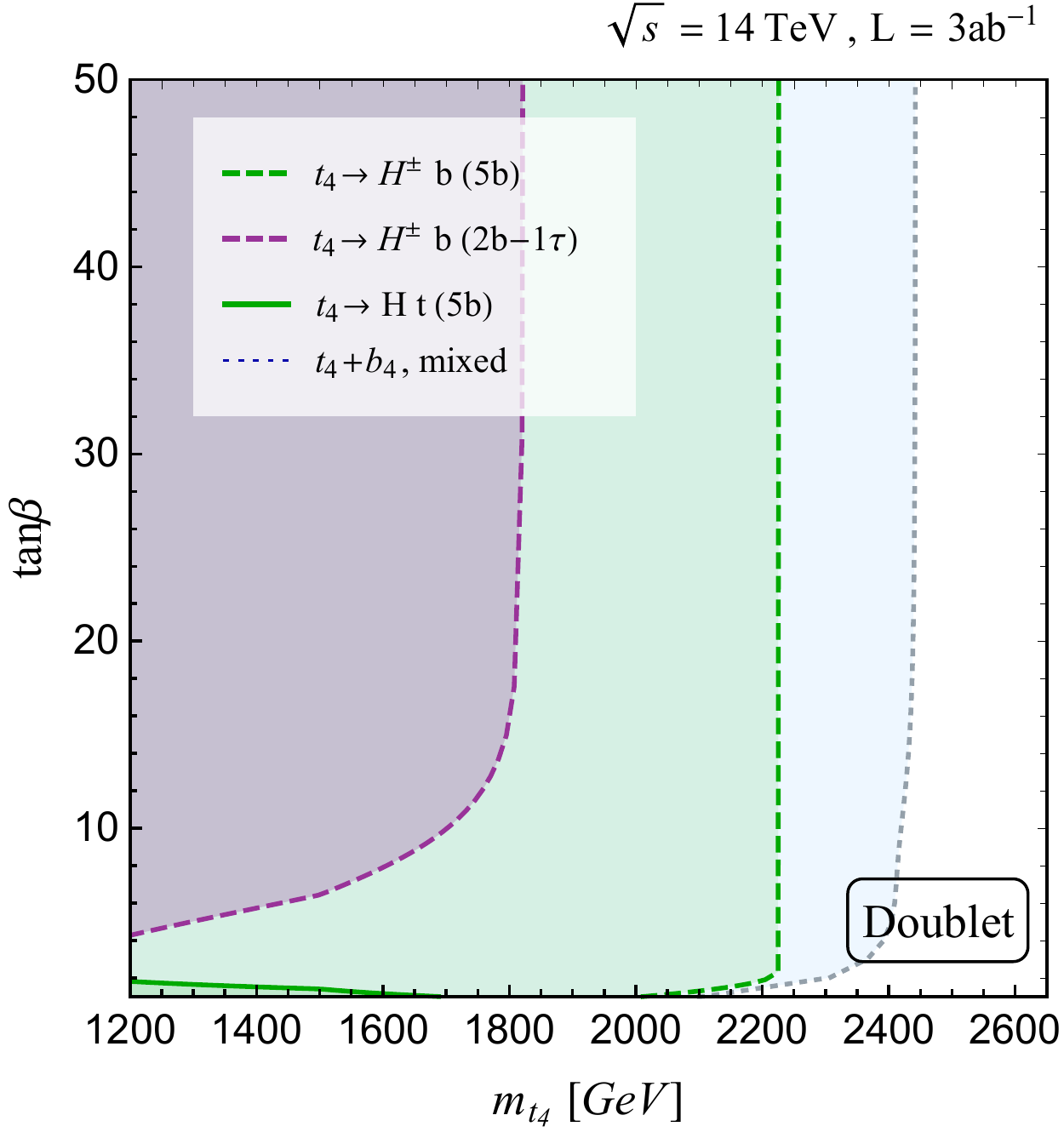}
\includegraphics[scale=0.5]{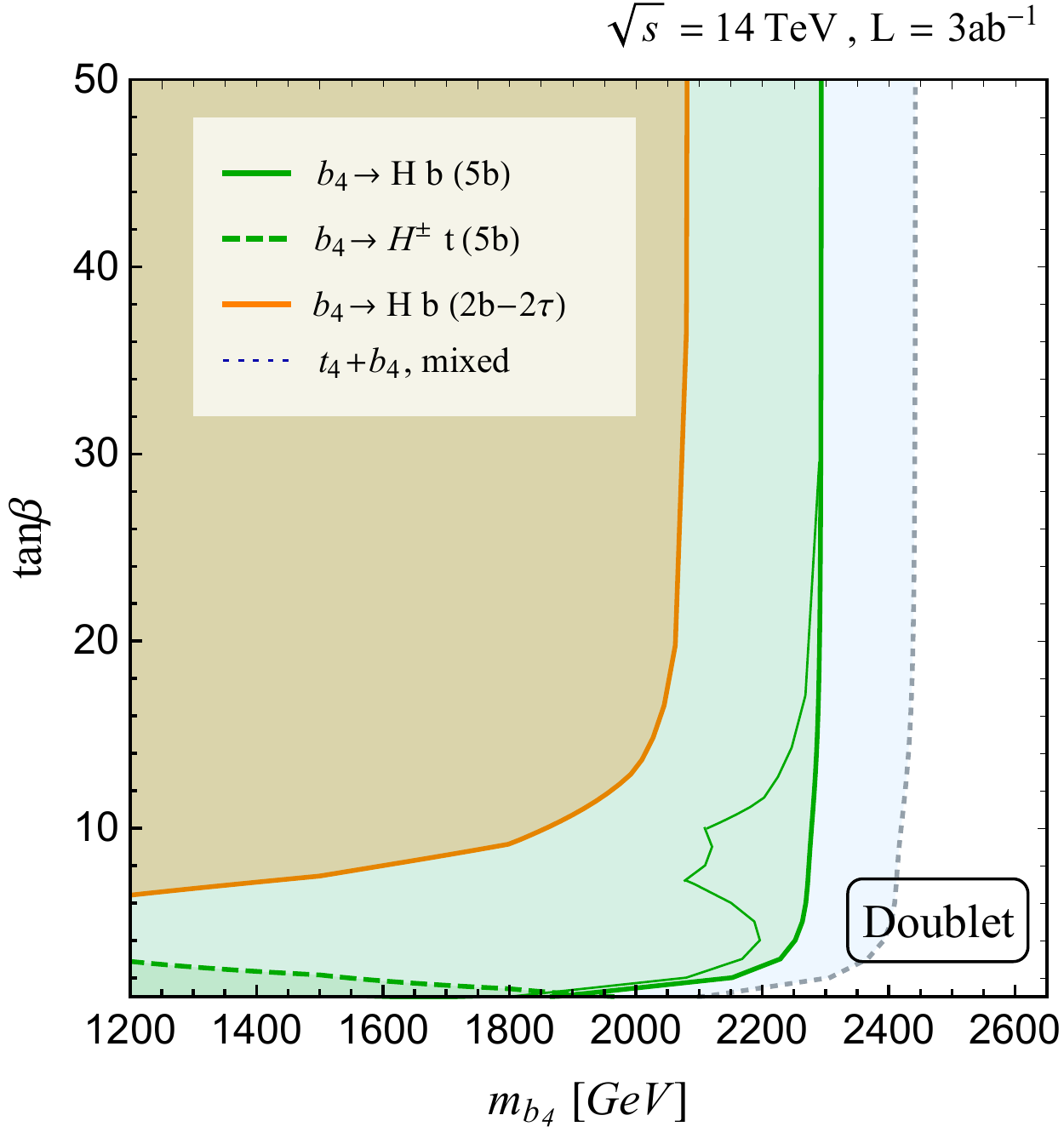}
\caption{95\% CL sensitivity regions for the HL-LHC in the $m_{t_{4}}$-$\tan\beta$ (left) and $m_{b_{4}}$-$\tan\beta$ (right) planes for singlet-like (top) and doublet-like (bottom) $t_{4}$ and $b_{4}$ in type-II 2HDM. The orange and purple regions follow from the upper limits obtained in Fig.~\ref{fig:BR_bounds}. The green shaded regions show the bounds on the $t_{4}/b_{4}\to H + t/b$ and $t_{4}/b_{4}\to H^{\pm}+b/t$ decays, respectively, using the 5b analysis of ref.~\cite{Dermisek:2020gbr}.  The blue region shows the combined reach for the $SU(2)$ doublet, assuming that both $t_{4}$ and $b_{4}$ are pair produced with the same mass and that decays to all heavy Higgses are kinematically open.
We show the envelope of the maximum reach among the three individual channels of $6b$, $4b2t$, and $2b4t$ in ref.~\cite{Dermisek:2020gbr} with the thin green line in the right panels.
}
\label{fig:tb_bounds}
\end{figure}

We recast model independent sensitivities obtained above as functions of vectorlike quark masses and $\tan\beta$  for the typical BRs of vectorlike quarks in type-II 2HDM discussed in section~\ref{sec:signal} and the usual BRs of heavy Higgses in this model. In figure
~\ref{fig:tb_bounds} we show the reach of the HL-LHC for singlet-like and doublet-like $t_4$ and $b_4$. The sensitivities to various decay modes based on $2b+2\tau$ and $2b+1\tau$ analyses are shown in orange and purple. 

For comparison, we also show the corresponding sensitivities obtained in ref.~\cite{Dermisek:2020gbr} in the green regions. To obtain the sensitivity to ${\rm BR}(b_{4}\to Hb)$ we have extended the analysis to account for cascade decays through ${\rm BR}(H\to t \bar{t})$. Thus, pair production of $b_{4}$ decaying to neutral Higgses can result in $6b$, $4b2t$, and $2b4t$ final states covering large, medium, and small $\tan\beta$ regions respectively. We show the envelope of the maximum reach among these three individual channels with the thin green line in figure~\ref{fig:tb_bounds} (right). It should be noted that all of these decays would simultaneously contribute to the $6b$ final state studied in ref.~\cite{Dermisek:2020gbr} (where we use the 5b analysis based on tagging at least 5 $b$-jets and top quarks are identified as $b$-jets through their decays to $W^{\pm}b$). Thus, including the regions of parameters where the events from each channel are combined results in the sensitivity to ${\rm BR}(b_{4}\to Hb)$ for masses of $b_4$ and $H$  up to 2.3  and 2.1 TeV, which is largely independent of $\tan\beta$ (solid green) except for small $\tan\beta\lesssim 4$ region. Similarly, the sensitivities to decays of $t_4$ and $b_4$ through the charged Higgs extends to vectorlike quark (Higgs) masses up to 2.2 (2) TeV which is also largely independent of $\tan\beta$.

When $t_{4}$ and $b_{4}$ are doublet-like, it is expected that they have almost the same masses and would be produced simultaneously, effectively doubling the production cross section of the signal. 
Allowing for all possible decay modes of the heavy Higgses, we see that the ${\rm BR}(b_{4}\to Hb)$ and  ${\rm BR} (t_4\to H^{\pm} b)$ decays provide the strongest reach. 
The reach of the signal when these channels are combined is shown in the light blue region and denoted by the mixed scenario in the legend. Thus, for doublet-like $t_{4}$ and $b_{4}$ the HL-LHC will have sensitivity to vectorlike quark (Higgs) masses up to 2.4 (2.2) TeV for $\tan\beta \gtrsim 4$. However, even for $\tan\beta \simeq 1$ this search strategy has sensitivity to vectorlike quark (Higgs) masses up to 2.1 (1.9) TeV.
\begin{figure}[t]
\centering
\includegraphics[scale=1]{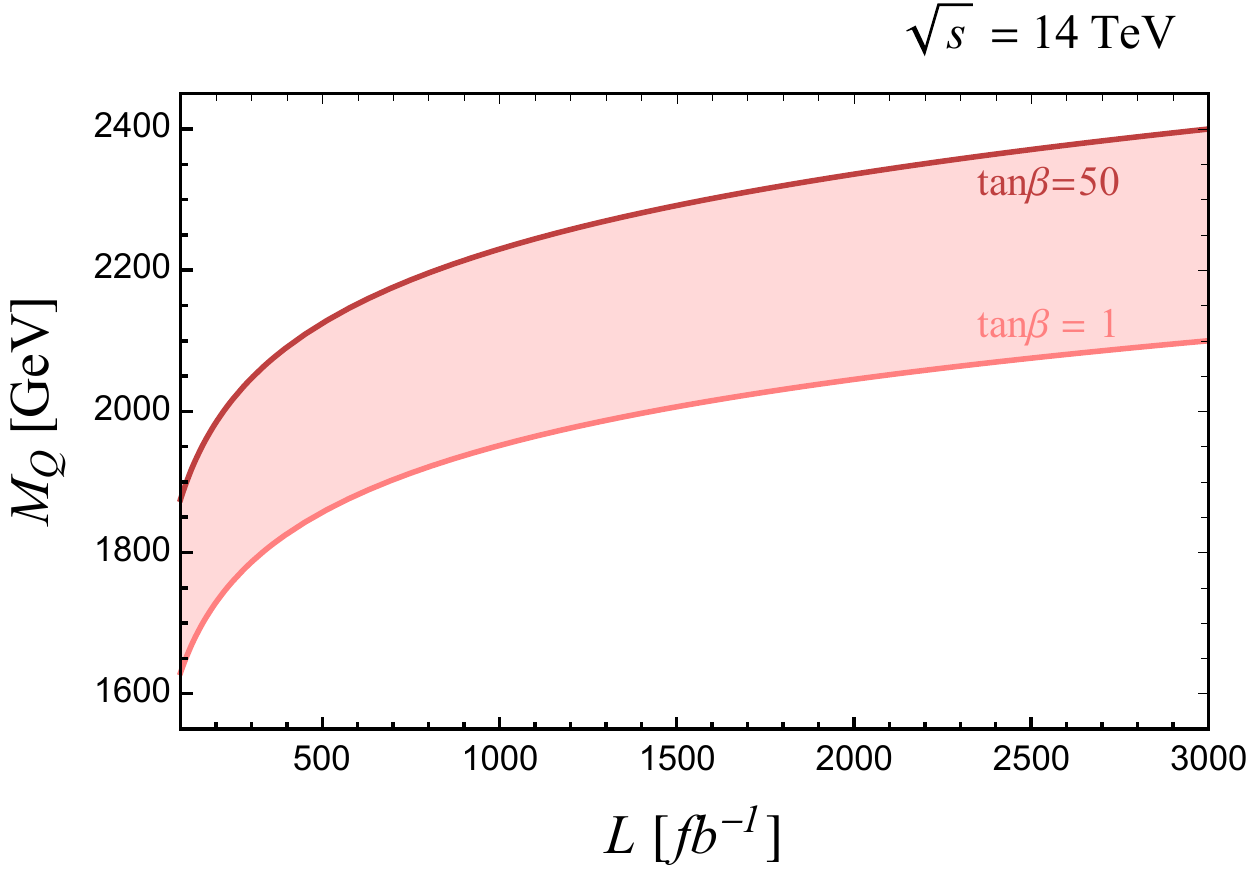}
\caption{95\% CL sensitivity to vectorlike $SU(2)$ doublet with mass $M_{Q}=m_{t_{4}}\simeq m_{b_{4}}$ decaying to heavy Higgses at the LHC running at 14 TeV with respect to the integrated luminosity. The band shows the ultimate reach based on the mixed scenario in Fig. \ref{fig:tb_bounds} between $\tan\beta = 1$ and $\tan\beta = 50$. }
\label{fig:MQ_vs_L}
\end{figure} 
In figure~\ref{fig:MQ_vs_L}, we use the sensitivities to the $SU(2)$ doublet quarks assuming decays through all heavy Higgses are kinematically open (the mixed $t_{4}+b_{4}$ scenario) to project the ultimate reach of the 14 TeV LHC to vectorlike quark masses at integrated luminosities between 100-3000 $fb^{-1}$. It is assumed that the signal and background efficiencies remain the same during the course of data collection. The two curves show the maximum reach based on our analysis for $\tan\beta=50$ and $\tan\beta=1$. We reiterate that these bounds imply sensitivity to heavy Higgs masses up to $m_{H}=M_{Q}-200$ GeV. It should also be noted that while search strategies using decays of heavy Higgses to $\tau$ leptons are generally weaker, due to dramatically smaller branching ratios ${\rm BR}(H / A \to \tau^+ \tau^-)\simeq 10\%$, the reach of the $2b+2\tau$ and $2b+1\tau$ analyses at $\tan\beta=50$ in the mixed scenario would be similar to the $\tan\beta=1$ curve in Fig.~\ref{fig:MQ_vs_L}. A striking result shown in  Fig.~\ref{fig:MQ_vs_L} is that even with currently available data, the discussed searches could lead to very strong limits on (or possibly the evidence of) vectorlike quarks and especially heavy Higgses for any $\tan\beta$.

\section{Conclusions}
\label{sec:conclusions}
In this paper, we presented search strategies for new quarks decaying through charged and neutral heavy Higgs bosons to final states with $\tau$ leptons, and multiple $b$-jets. In extensions of type-II 2HDM, the decays of vectorlike quarks to heavy Higgses can dominate over those to SM bosons, and thus dramatically alter the projected limits on the masses of new quarks. Likewise, the production of heavy Higgses from decays of vectorlike quarks leads to QCD-size rates for signals of both heavy charged and neutral Higgs bosons. We motivate, in particular, final states with two or one $\tau$ tagged jets based on the pattern of branching ratios of heavy Higgses in type-II 2HDM. This strategy is simultaneously sensitive to multiple possible decay chains such as $b_{4}\to Hb$, $b_{4}\to H^\pm t$, and $t_{4}\to H^\pm b$. 

Decay modes of heavy Higgses to $\tau$ leptons typically offer much cleaner signatures at the LHC despite much weaker rates compared to, e.g. $H(A)\to b\bar{b}$. As our analyses are also based on tagging multiple $b$-jets, QCD-driven events, such as $tt(jj)$, comprise a formidable source of SM background. However, the kinematic advantage of the heavy resonance in the decay chain allows to suppress the dominant backgrounds due to markedly different correlations between $\tau$- and $b$-tagged jets. These search strategies would also be applicable to many models of new physics in which new quarks are pair produced at the LHC decaying to heavy bosons with couplings to third-generation leptons. We therefore encourage our experimental colleagues to search for the $\tau$ signatures with high $H_T$.

The model-independent upper limits we find on the $b_{4}\to Hb$, $b_{4}\to H^\pm t$, and $t_{4}\to H^\pm b$ decays show that searches focusing on $\tau$ leptons can probe vectorlike quark masses up to 1.8 TeV and 2 TeV at the HL-LHC when decaying to charged and neutral heavy Higgses, respectively. These limits imply sensitivity to charged and neutral Higgs masses to 1.6 TeV and 1.8 TeV. We find that these channels offer only comparable reach with respect to searches based only on $b$-tagging~\cite{Dermisek:2020gbr}. However, note that the utility of independent search channels would offer further insight to the heavy Higgs sector in the event of a discovery. In particular, while our model-independent limits are motivated by type-II 2HDM scenarios where heavy Higgses decay mostly to quarks, other 2HDM variants (or scenarios with heavy charged or neutral vector bosons) would include different hierarchies of branching ratios. The observation of vectorlike quarks in the absence of either $\tau$- or only $b$-tagged channels could be used to disentangle the underlying dynamics of the Higgs sector itself.

Among our results, we recast the model independent bounds explored in this paper to sensitivities of vectorlike singlet and doublet quark masses in type-II 2HDM at the HL-LHC. We also present the corresponding sensitivity of the 5b analyses presented in ref~\cite{Dermisek:2020gbr} based on comprehensive searches of a $6b$ final state. Further, we extend the 5b analyses to include all sizable decay modes of charged and neutral heavy Higgses. For singlet-like quarks we find that the combined sensitivity to vectorlike quark and heavy Higgs masses extends up to 2.2 TeV and 2 TeV, respectively. The limits are improved when new quarks are doublet-like as both top- and bottom-like quarks are expected to have similar masses which would effectively double the rates we consider. In this case, we find sensitivity of the HL-LHC up to 2.4 (2.2) TeV for vectorlike quark (Higgs) masses.

We remark that the limits we obtain for new quark and heavy Higgs masses in the type-II 2HDM, both for singlet- and doublet-like $b_{4}$ quarks, are largely independent of $\tan\beta$, a striking difference of the projected limits compared to those of searches for heavy Higgses without vectorlike quarks. This is mirrored also in the limits for doublet-like $t_{4}$. Further, when the production of doublet-like $t_4$ and $b_4$ are combined, even the weakest sensitivity we obtain, $M_{Q}\simeq 2.1$ TeV for $\tan\beta=1$, is still quite strong compared to existing searches. This implies robust sensitivities in searches for new quarks and heavy Higgs bosons, even with current data.

\acknowledgments
The work of RD was supported in part by the U.S. Department of Energy under grant number {DE}-SC0010120. NM acknowledges partial support by the U.S. Department of Energy under contracts No. DEAC02-06CH11357 while at
Argonne National Laboratory. TRIUMF receives federal funding via a contribution agreement with the National Research Council of Canada. SS acknowledges support from the National Research Foundation of Korea (NRF-2020R1I1A3072747).

\appendix

\section{Poisson statistics}
\label{app:poisson}
In this appendix we review our methodology to obtain upper limits on signal cross sections based on Poisson statistics. For the analyses based on a single final state, we derive an upper limit on the number of signal events when the expected number of background events, $b$, is known via~\cite{Tanabashi:2018oca}
\begin{align}
s_{\rm up} (\alpha, n, b) &= \frac{1}{2} F_{\chi^2}^{-1} [p, 2 (n+1)] -b \\
p &= 1- \alpha (1 - F_{\chi^2} [2b, 2(n+1)]) ,
\end{align}
where $n$ is the total number of events observed, $F_{\chi^2}$ is the $\chi^2$ cumulative distribution, and the confidence level of the resulting upper bound is given by $(1-\alpha)$. In the absence of signal events, the median of upper limits can be calculated by sampling over the background distribution
\begin{align}
\bar s_{\rm up} (\alpha, b) = s_{\rm up} (\alpha, b, b) \; .
\end{align}
The upper limit on the signal cross section is then given by
\begin{align}
\sigma_{\rm sig}^{\rm up} = \frac{ \bar s_{\rm up} (\alpha, \sigma_B {\cal L})}{{\cal L} \cdot \varepsilon_{\rm sig}},
\end{align}
where ${\cal L}$ is the integrated luminosity and $\varepsilon_{\rm sig}$ is the signal efficiency, optimized over the chosen kinematic cuts. We compare the derived $\sigma_{\rm sig}^{\rm up}$ to our signal cross sections which are fixed for a given mass of $t_{4}$ or $b_{4}$ and corresponding branching ratios.

In our main results, we have calculated the statistical combination of vectorlike quark cascade decays contributing independently to the $2b+2\tau$ and $2b+1\tau$ final states. In general, assuming $X$ independent final states (e.g. resulting from multiple decay channels of some particle(s)) the total probability of observing $N = \sum_{k=1}^{X} n_{k}$ events, where $n_{k}$ is the number of observed events in the $k$-th channel, given an expected number of background events $b_{k}$ for each final state is given by
\begin{align}
\prod_{k=1}^{X}p(n_{k}, s_{k} + b_{k}),
\end{align}
where $s_{k}$ is the expected number of signal events in the $k$-th channel. Following the same reasoning as in the case of a single channel, the upper limit on the signal cross section can be derived by calculating the confidence interval for $X$ independent poisson distributions~\cite{Cowan:1998ji}
\begin{align}
(1-\alpha)=\frac{\int_{0}^{\sigma_{sig}^{up}}d\sigma \prod_{k=1}^{X}p(b_{k}, s_k + b_{k})}{\int_{0}^{\infty}d\sigma \prod_{k=1}^{X}p(b_{k}, s_k + b_{k})},
\end{align}
where $b_k = \sigma_{B_k} {\cal L}$, $s_k = \epsilon_{k}\sigma {\cal L}$ and $\epsilon_{k}$ is the signal efficiency in the $k$-th channel. We compute the upper limit on the cross section, $\sigma_{sig}^{up}$, numerically and compare this to the signal for a given vectorlike quark mass and cascade branching ratios through heavy Higgses.

\bibliography{VLQ_taus}

\providecommand{\href}[2]{#2}\begingroup\raggedright\begin{thebibliography}{10}

\bibitem{Dermisek:2019vkc}
R.~Derm\'\i{}\v{s}ek, E.~Lunghi and S.~Shin, \emph{{Hunting for Vectorlike
  Quarks}}, \href{http://dx.doi.org/10.1007/JHEP04(2019)019}{\emph{JHEP} {\bf
  04} (2019) 019}, [\href{http://arxiv.org/abs/1901.03709}{{\tt 1901.03709}}].

\bibitem{Dermisek:2020gbr}
R.~Dermisek, E.~Lunghi, N.~McGinnis and S.~Shin, \emph{{Signals with six bottom
  quarks for charged and neutral Higgs bosons}},
  \href{http://dx.doi.org/10.1007/JHEP07(2020)241}{\emph{JHEP} {\bf 07} (2020)
  241}, [\href{http://arxiv.org/abs/2005.07222}{{\tt 2005.07222}}].

\bibitem{Aad:2020zxo}
{\scshape ATLAS} collaboration, G.~Aad et~al., \emph{{Search for heavy Higgs
  bosons decaying into two tau leptons with the ATLAS detector using $pp$
  collisions at $\sqrt{s}=13$ TeV}},
  \href{http://dx.doi.org/10.1103/PhysRevLett.125.051801}{\emph{Phys. Rev.
  Lett.} {\bf 125} (2020) 051801}, [\href{http://arxiv.org/abs/2002.12223}{{\tt
  2002.12223}}].

\bibitem{Sirunyan:2018zut}
{\scshape CMS} collaboration, A.~M. Sirunyan et~al., \emph{{Search for
  additional neutral MSSM Higgs bosons in the $\tau\tau$ final state in
  proton-proton collisions at $\sqrt{s}=$ 13 TeV}},
  \href{http://dx.doi.org/10.1007/JHEP09(2018)007}{\emph{JHEP} {\bf 09} (2018)
  007}, [\href{http://arxiv.org/abs/1803.06553}{{\tt 1803.06553}}].

\bibitem{Aaboud:2018gjj}
{\scshape ATLAS} collaboration, M.~Aaboud et~al., \emph{{Search for charged
  Higgs bosons decaying via $H^{\pm} \to \tau^{\pm}\nu_{\tau}$ in the
  $\tau$+jets and $\tau$+lepton final states with 36 fb$^{-1}$ of $pp$
  collision data recorded at $\sqrt{s} = 13$ TeV with the ATLAS experiment}},
  \href{http://dx.doi.org/10.1007/JHEP09(2018)139}{\emph{JHEP} {\bf 09} (2018)
  139}, [\href{http://arxiv.org/abs/1807.07915}{{\tt 1807.07915}}].

\bibitem{Sirunyan:2019hkq}
{\scshape CMS} collaboration, A.~M. Sirunyan et~al., \emph{{Search for charged
  Higgs bosons in the H$^{\pm}$ $\to$ $\tau^{\pm}\nu_\tau$ decay channel in
  proton-proton collisions at $\sqrt{s} =$ 13 TeV}},
  \href{http://dx.doi.org/10.1007/JHEP07(2019)142}{\emph{JHEP} {\bf 07} (2019)
  142}, [\href{http://arxiv.org/abs/1903.04560}{{\tt 1903.04560}}].

\bibitem{Aaboud:2018cwk}
{\scshape ATLAS} collaboration, M.~Aaboud et~al., \emph{{Search for charged
  Higgs bosons decaying into top and bottom quarks at $\sqrt{s}$ = 13 TeV with
  the ATLAS detector}},
  \href{http://dx.doi.org/10.1007/JHEP11(2018)085}{\emph{JHEP} {\bf 11} (2018)
  085}, [\href{http://arxiv.org/abs/1808.03599}{{\tt 1808.03599}}].

\bibitem{Sirunyan:2019arl}
{\scshape CMS} collaboration, A.~M. Sirunyan et~al., \emph{{Search for a
  charged Higgs boson decaying into top and bottom quarks in events with
  electrons or muons in proton-proton collisions at $ \sqrt{\mathrm{s}} $ = 13
  TeV}}, \href{http://dx.doi.org/10.1007/JHEP01(2020)096}{\emph{JHEP} {\bf 01}
  (2020) 096}, [\href{http://arxiv.org/abs/1908.09206}{{\tt 1908.09206}}].

\bibitem{ATLAS:2020jqj}
{\scshape ATLAS} collaboration, \emph{{Search for charged Higgs bosons decaying
  into a top-quark and a bottom-quark at $\sqrt s$ = 13 TeV with the ATLAS
  detector}}, .

\bibitem{Sirunyan:2018taj}
{\scshape CMS} collaboration, A.~M. Sirunyan et~al., \emph{{Search for beyond
  the standard model Higgs bosons decaying into a $\mathrm{b\overline{b}}$ pair
  in pp collisions at $\sqrt{s} =$ 13 TeV}},
  \href{http://dx.doi.org/10.1007/JHEP08(2018)113}{\emph{JHEP} {\bf 08} (2018)
  113}, [\href{http://arxiv.org/abs/1805.12191}{{\tt 1805.12191}}].

\bibitem{Aad:2019zwb}
{\scshape ATLAS} collaboration, G.~Aad et~al., \emph{{Search for heavy neutral
  Higgs bosons produced in association with $b$-quarks and decaying into
  $b$-quarks at $\sqrt{s}=13$ TeV with the ATLAS detector}},
  \href{http://dx.doi.org/10.1103/PhysRevD.102.032004}{\emph{Phys. Rev. D} {\bf
  102} (2020) 032004}, [\href{http://arxiv.org/abs/1907.02749}{{\tt
  1907.02749}}].

\bibitem{Dermisek:2011xu}
R.~Dermisek, S.-G. Kim and A.~Raval, \emph{{New Vector Boson Near the Z-pole
  and the Puzzle in Precision Electroweak Data}},
  \href{http://dx.doi.org/10.1103/PhysRevD.84.035006}{\emph{Phys. Rev. D} {\bf
  84} (2011) 035006}, [\href{http://arxiv.org/abs/1105.0773}{{\tt 1105.0773}}].

\bibitem{Dermisek:2012qx}
R.~Dermisek, S.-G. Kim and A.~Raval, \emph{{Z' near the Z-pole}},
  \href{http://dx.doi.org/10.1103/PhysRevD.85.075022}{\emph{Phys. Rev. D} {\bf
  85} (2012) 075022}, [\href{http://arxiv.org/abs/1201.0315}{{\tt 1201.0315}}].

\bibitem{Kawamura:2019rth}
J.~Kawamura, S.~Raby and A.~Trautner, \emph{{Complete vectorlike fourth family
  and new U(1)' for muon anomalies}},
  \href{http://dx.doi.org/10.1103/PhysRevD.100.055030}{\emph{Phys. Rev. D} {\bf
  100} (2019) 055030}, [\href{http://arxiv.org/abs/1906.11297}{{\tt
  1906.11297}}].

\bibitem{Kawamura:2019hxp}
J.~Kawamura, S.~Raby and A.~Trautner, \emph{{Complete vectorlike fourth family
  with U(1)' : A global analysis}},
  \href{http://dx.doi.org/10.1103/PhysRevD.101.035026}{\emph{Phys. Rev. D} {\bf
  101} (2020) 035026}, [\href{http://arxiv.org/abs/1911.11075}{{\tt
  1911.11075}}].

\bibitem{Dermisek:2019heo}
R.~Dermisek, E.~Lunghi and S.~Shin, \emph{{Cascade decays of heavy Higgs bosons
  through vectorlike quarks in two Higgs doublet models}},
  \href{http://dx.doi.org/10.1007/JHEP03(2020)029}{\emph{JHEP} {\bf 03} (2020)
  029}, [\href{http://arxiv.org/abs/1907.07188}{{\tt 1907.07188}}].

\bibitem{Dermisek:2015oja}
R.~Dermisek, E.~Lunghi and S.~Shin, \emph{{Two Higgs doublet model with
  vectorlike leptons and contributions to $pp\to WW$ and $H\to WW$}},
  \href{http://dx.doi.org/10.1007/JHEP02(2016)119}{\emph{JHEP} {\bf 02} (2016)
  119}, [\href{http://arxiv.org/abs/1509.04292}{{\tt 1509.04292}}].

\bibitem{Dermisek:2015vra}
R.~Dermisek, E.~Lunghi and S.~Shin, \emph{{Contributions of flavor violating
  couplings of a Higgs boson to $pp \to WW$}},
  \href{http://dx.doi.org/10.1007/JHEP08(2015)126}{\emph{JHEP} {\bf 08} (2015)
  126}, [\href{http://arxiv.org/abs/1503.08829}{{\tt 1503.08829}}].

\bibitem{Dermisek:2015hue}
R.~Dermisek, E.~Lunghi and S.~Shin, \emph{{New decay modes of heavy Higgs
  bosons in a two Higgs doublet model with vectorlike leptons}},
  \href{http://dx.doi.org/10.1007/JHEP05(2016)148}{\emph{JHEP} {\bf 05} (2016)
  148}, [\href{http://arxiv.org/abs/1512.07837}{{\tt 1512.07837}}].

\bibitem{Dermisek:2016via}
R.~Dermisek, E.~Lunghi and S.~Shin, \emph{{New constraints and discovery
  potential for Higgs to Higgs cascade decays through vectorlike leptons}},
  \href{http://dx.doi.org/10.1007/JHEP10(2016)081}{\emph{JHEP} {\bf 10} (2016)
  081}, [\href{http://arxiv.org/abs/1608.00662}{{\tt 1608.00662}}].

\bibitem{CidVidal:2018eel}
X.~Cid~Vidal et~al., \emph{{Report from Working Group 3}: {Beyond the Standard
  Model physics at the HL-LHC and HE-LHC}},
  \href{http://dx.doi.org/10.23731/CYRM-2019-007.585}{\emph{CERN Yellow Rep.
  Monogr.} {\bf 7} (2019) 585--865},
  [\href{http://arxiv.org/abs/1812.07831}{{\tt 1812.07831}}].

\bibitem{Dermisek:2012as}
R.~Dermisek, \emph{{Insensitive Unification of Gauge Couplings}},
  \href{http://dx.doi.org/10.1016/j.physletb.2012.06.037}{\emph{Phys. Lett. B}
  {\bf 713} (2012) 469--472}, [\href{http://arxiv.org/abs/1204.6533}{{\tt
  1204.6533}}].

\bibitem{Dermisek:2012ke}
R.~Dermisek, \emph{{Unification of gauge couplings in the standard model with
  extra vectorlike families}},
  \href{http://dx.doi.org/10.1103/PhysRevD.87.055008}{\emph{Phys. Rev. D} {\bf
  87} (2013) 055008}, [\href{http://arxiv.org/abs/1212.3035}{{\tt 1212.3035}}].

\bibitem{Dermisek:2017ihj}
R.~Dermisek and N.~McGinnis, \emph{{Mass scale of vectorlike matter and
  superpartners from IR fixed point predictions of gauge and top Yukawa
  couplings}}, \href{http://dx.doi.org/10.1103/PhysRevD.97.055009}{\emph{Phys.
  Rev. D} {\bf 97} (2018) 055009}, [\href{http://arxiv.org/abs/1712.03527}{{\tt
  1712.03527}}].

\bibitem{Dermisek:2018hxq}
R.~Derm\'\i{}\v{s}ek and N.~McGinnis, \emph{{Top-bottom-tau Yukawa coupling
  unification in the MSSM plus one vectorlike family and fermion masses as IR
  fixed points}},
  \href{http://dx.doi.org/10.1103/PhysRevD.99.035033}{\emph{Phys. Rev. D} {\bf
  99} (2019) 035033}, [\href{http://arxiv.org/abs/1810.12474}{{\tt
  1810.12474}}].

\bibitem{Dermisek:2018ujw}
R.~Derm\'\i{}\v{s}ek and N.~McGinnis, \emph{{Seven largest couplings of the
  standard model as IR fixed points}},
  \href{http://dx.doi.org/10.1103/PhysRevLett.122.181803}{\emph{Phys. Rev.
  Lett.} {\bf 122} (2019) 181803}, [\href{http://arxiv.org/abs/1812.05240}{{\tt
  1812.05240}}].

\bibitem{Dermisek:2016tzw}
R.~Dermisek, \emph{{Loop suppressed electroweak symmetry breaking and naturally
  heavy superpartners}},
  \href{http://dx.doi.org/10.1103/PhysRevD.95.015002}{\emph{Phys. Rev. D} {\bf
  95} (2017) 015002}, [\href{http://arxiv.org/abs/1606.09031}{{\tt
  1606.09031}}].

\bibitem{Cohen:2020ohi}
T.~Cohen, N.~Craig, S.~Koren, M.~Mccullough and J.~Tooby-Smith,
  \emph{{Supersoft Top Squarks}},
  \href{http://dx.doi.org/10.1103/PhysRevLett.125.151801}{\emph{Phys. Rev.
  Lett.} {\bf 125} (2020) 151801}, [\href{http://arxiv.org/abs/2002.12630}{{\tt
  2002.12630}}].

\bibitem{Dermisek:2020cod}
R.~Dermisek, K.~Hermanek and N.~McGinnis, \emph{{Highly Enhanced Contributions
  of Heavy Higgs Bosons and New Leptons to Muon $g$\ensuremath{-}2 and
  Prospects at Future Colliders}},
  \href{http://dx.doi.org/10.1103/PhysRevLett.126.191801}{\emph{Phys. Rev.
  Lett.} {\bf 126} (2021) 191801}, [\href{http://arxiv.org/abs/2011.11812}{{\tt
  2011.11812}}].

\bibitem{Dermisek:2021ajd}
R.~Dermisek, K.~Hermanek and N.~McGinnis, \emph{{Muon $g-2$ in two Higgs
  doublet models with vectorlike leptons}},
  \href{http://arxiv.org/abs/2103.05645}{{\tt 2103.05645}}.

\bibitem{Dermisek:2014cia}
R.~Dermisek, A.~Raval and S.~Shin, \emph{{Effects of vectorlike leptons on
  $h\to 4\ell$ and the connection to the muon g-2 anomaly}},
  \href{http://dx.doi.org/10.1103/PhysRevD.90.034023}{\emph{Phys. Rev. D} {\bf
  90} (2014) 034023}, [\href{http://arxiv.org/abs/1406.7018}{{\tt 1406.7018}}].

\bibitem{Dermisek:2013gta}
R.~Dermisek and A.~Raval, \emph{{Explanation of the Muon g-2 Anomaly with
  Vectorlike Leptons and its Implications for Higgs Decays}},
  \href{http://dx.doi.org/10.1103/PhysRevD.88.013017}{\emph{Phys. Rev. D} {\bf
  88} (2013) 013017}, [\href{http://arxiv.org/abs/1305.3522}{{\tt 1305.3522}}].

\bibitem{Czarnecki:2001pv}
A.~Czarnecki and W.~J. Marciano, \emph{{The Muon anomalous magnetic moment: A
  Harbinger for 'new physics'}},
  \href{http://dx.doi.org/10.1103/PhysRevD.64.013014}{\emph{Phys. Rev. D} {\bf
  64} (2001) 013014}, [\href{http://arxiv.org/abs/hep-ph/0102122}{{\tt
  hep-ph/0102122}}].

\bibitem{Kannike:2011ng}
K.~Kannike, M.~Raidal, D.~M. Straub and A.~Strumia, \emph{{Anthropic solution
  to the magnetic muon anomaly: the charged see-saw}},
  \href{http://dx.doi.org/10.1007/JHEP02(2012)106}{\emph{JHEP} {\bf 02} (2012)
  106}, [\href{http://arxiv.org/abs/1111.2551}{{\tt 1111.2551}}].

\bibitem{Dermisek:2014qca}
R.~Dermisek, J.~P. Hall, E.~Lunghi and S.~Shin, \emph{{Limits on Vectorlike
  Leptons from Searches for Anomalous Production of Multi-Lepton Events}},
  \href{http://dx.doi.org/10.1007/JHEP12(2014)013}{\emph{JHEP} {\bf 12} (2014)
  013}, [\href{http://arxiv.org/abs/1408.3123}{{\tt 1408.3123}}].

\bibitem{Kumar:2015tna}
N.~Kumar and S.~P. Martin, \emph{{Vectorlike Leptons at the Large Hadron
  Collider}}, \href{http://dx.doi.org/10.1103/PhysRevD.92.115018}{\emph{Phys.
  Rev. D} {\bf 92} (2015) 115018}, [\href{http://arxiv.org/abs/1510.03456}{{\tt
  1510.03456}}].

\bibitem{Bhattiprolu:2019vdu}
P.~N. Bhattiprolu and S.~P. Martin, \emph{{Prospects for vectorlike leptons at
  future proton-proton colliders}},
  \href{http://dx.doi.org/10.1103/PhysRevD.100.015033}{\emph{Phys. Rev. D} {\bf
  100} (2019) 015033}, [\href{http://arxiv.org/abs/1905.00498}{{\tt
  1905.00498}}].

\bibitem{Freitas:2020ttd}
F.~F. Freitas, J.~a. Gon\c{c}alves, A.~P. Morais and R.~Pasechnik,
  \emph{{Phenomenology of vector-like leptons with Deep Learning at the Large
  Hadron Collider}},
  \href{http://dx.doi.org/10.1007/JHEP01(2021)076}{\emph{JHEP} {\bf 01} (2021)
  076}, [\href{http://arxiv.org/abs/2010.01307}{{\tt 2010.01307}}].

\bibitem{Bissmann:2020lge}
S.~Bi\ss{}mann, G.~Hiller, C.~Hormigos-Feliu and D.~F. Litim,
  \emph{{Multi-lepton signatures of vector-like leptons with flavor}},
  \href{http://dx.doi.org/10.1140/epjc/s10052-021-08886-3}{\emph{Eur. Phys. J.
  C} {\bf 81} (2021) 101}, [\href{http://arxiv.org/abs/2011.12964}{{\tt
  2011.12964}}].

\bibitem{Kawamura:2021ygg}
J.~Kawamura and S.~Raby, \emph{{Signal of four muons or more from a vector-like
  lepton decaying to a muon-philic Z' boson at the LHC}},
  \href{http://dx.doi.org/10.1103/PhysRevD.104.035007}{\emph{Phys. Rev. D} {\bf
  104} (2021) 035007}, [\href{http://arxiv.org/abs/2104.04461}{{\tt
  2104.04461}}].

\bibitem{Choudhury:2021nib}
D.~Choudhury, K.~Deka and N.~Kumar, \emph{{Looking for a vectorlike B quark at
  the LHC using jet substructure}},
  \href{http://dx.doi.org/10.1103/PhysRevD.104.035004}{\emph{Phys. Rev. D} {\bf
  104} (2021) 035004}, [\href{http://arxiv.org/abs/2103.10655}{{\tt
  2103.10655}}].

\bibitem{Raby:2017igl}
S.~Raby and A.~Trautner, \emph{{Vectorlike chiral fourth family to explain muon
  anomalies}}, \href{http://dx.doi.org/10.1103/PhysRevD.97.095006}{\emph{Phys.
  Rev. D} {\bf 97} (2018) 095006}, [\href{http://arxiv.org/abs/1712.09360}{{\tt
  1712.09360}}].

\bibitem{Crivellin:2018qmi}
A.~Crivellin, M.~Hoferichter and P.~Schmidt-Wellenburg, \emph{{Combined
  explanations of $(g-2)_{\mu,e}$ and implications for a large muon EDM}},
  \href{http://dx.doi.org/10.1103/PhysRevD.98.113002}{\emph{Phys. Rev. D} {\bf
  98} (2018) 113002}, [\href{http://arxiv.org/abs/1807.11484}{{\tt
  1807.11484}}].

\bibitem{Barman:2018jhz}
B.~Barman, D.~Borah, L.~Mukherjee and S.~Nandi, \emph{{Correlating the
  anomalous results in $b \to s$ decays with inert Higgs doublet dark matter
  and muon $(g-2)$}},
  \href{http://dx.doi.org/10.1103/PhysRevD.100.115010}{\emph{Phys. Rev. D} {\bf
  100} (2019) 115010}, [\href{http://arxiv.org/abs/1808.06639}{{\tt
  1808.06639}}].

\bibitem{Arnan:2019uhr}
P.~Arnan, A.~Crivellin, M.~Fedele and F.~Mescia, \emph{{Generic Loop Effects of
  New Scalars and Fermions in $b\to s\ell^+\ell^-$, $(g-2)_\mu$ and a
  Vector-like $4^{\rm th}$ Generation}},
  \href{http://dx.doi.org/10.1007/JHEP06(2019)118}{\emph{JHEP} {\bf 06} (2019)
  118}, [\href{http://arxiv.org/abs/1904.05890}{{\tt 1904.05890}}].

\bibitem{Endo:2020tkb}
M.~Endo and S.~Mishima, \emph{{Muon $g − 2$ and CKM unitarity in extra lepton
  models}}, \href{http://dx.doi.org/10.1007/JHEP08(2020)004}{\emph{JHEP} {\bf
  08} (2020) 004}, [\href{http://arxiv.org/abs/2005.03933}{{\tt 2005.03933}}].

\bibitem{Crivellin:2020ebi}
A.~Crivellin, F.~Kirk, C.~A. Manzari and M.~Montull, \emph{{Global Electroweak
  Fit and Vector-Like Leptons in Light of the Cabibbo Angle Anomaly}},
  \href{http://dx.doi.org/10.1007/JHEP12(2020)166}{\emph{JHEP} {\bf 12} (2020)
  166}, [\href{http://arxiv.org/abs/2008.01113}{{\tt 2008.01113}}].

\bibitem{Lu:2017uur}
Q.~Lu, D.~E. Morrissey and A.~M. Wijangco, \emph{{Higgs Boson Decays to Dark
  Photons through the Vectorized Lepton Portal}},
  \href{http://dx.doi.org/10.1007/JHEP06(2017)138}{\emph{JHEP} {\bf 06} (2017)
  138}, [\href{http://arxiv.org/abs/1705.08896}{{\tt 1705.08896}}].

\bibitem{Kowalska:2017iqv}
K.~Kowalska and E.~M. Sessolo, \emph{{Expectations for the muon g-2 in
  simplified models with dark matter}},
  \href{http://dx.doi.org/10.1007/JHEP09(2017)112}{\emph{JHEP} {\bf 09} (2017)
  112}, [\href{http://arxiv.org/abs/1707.00753}{{\tt 1707.00753}}].

\bibitem{Calibbi:2018rzv}
L.~Calibbi, R.~Ziegler and J.~Zupan, \emph{{Minimal models for dark matter and
  the muon g\ensuremath{-}2 anomaly}},
  \href{http://dx.doi.org/10.1007/JHEP07(2018)046}{\emph{JHEP} {\bf 07} (2018)
  046}, [\href{http://arxiv.org/abs/1804.00009}{{\tt 1804.00009}}].

\bibitem{Jana:2020joi}
S.~Jana, P.~K. Vishnu, W.~Rodejohann and S.~Saad, \emph{{Dark matter assisted
  lepton anomalous magnetic moments and neutrino masses}},
  \href{http://dx.doi.org/10.1103/PhysRevD.102.075003}{\emph{Phys. Rev. D} {\bf
  102} (2020) 075003}, [\href{http://arxiv.org/abs/2008.02377}{{\tt
  2008.02377}}].

\bibitem{Alwall:2014hca}
J.~Alwall, R.~Frederix, S.~Frixione, V.~Hirschi, F.~Maltoni, O.~Mattelaer
  et~al., \emph{{The automated computation of tree-level and next-to-leading
  order differential cross sections, and their matching to parton shower
  simulations}}, \href{http://dx.doi.org/10.1007/JHEP07(2014)079}{\emph{JHEP}
  {\bf 07} (2014) 079}, [\href{http://arxiv.org/abs/1405.0301}{{\tt
  1405.0301}}].

\bibitem{Czakon:2011xx}
M.~Czakon and A.~Mitov, \emph{{Top++: A Program for the Calculation of the
  Top-Pair Cross-Section at Hadron Colliders}},
  \href{http://dx.doi.org/10.1016/j.cpc.2014.06.021}{\emph{Comput. Phys.
  Commun.} {\bf 185} (2014) 2930}, [\href{http://arxiv.org/abs/1112.5675}{{\tt
  1112.5675}}].

\bibitem{Degrande:2011ua}
C.~Degrande, C.~Duhr, B.~Fuks, D.~Grellscheid, O.~Mattelaer and T.~Reiter,
  \emph{{UFO - The Universal FeynRules Output}},
  \href{http://dx.doi.org/10.1016/j.cpc.2012.01.022}{\emph{Comput. Phys.
  Commun.} {\bf 183} (2012) 1201--1214},
  [\href{http://arxiv.org/abs/1108.2040}{{\tt 1108.2040}}].

\bibitem{Sjostrand:2006za}
T.~Sjostrand, S.~Mrenna and P.~Z. Skands, \emph{{PYTHIA 6.4 Physics and
  Manual}}, \href{http://dx.doi.org/10.1088/1126-6708/2006/05/026}{\emph{JHEP}
  {\bf 05} (2006) 026}, [\href{http://arxiv.org/abs/hep-ph/0603175}{{\tt
  hep-ph/0603175}}].

\bibitem{Sjostrand:2014zea}
T.~Sj\"ostrand, S.~Ask, J.~R. Christiansen, R.~Corke, N.~Desai, P.~Ilten
  et~al., \emph{{An introduction to PYTHIA 8.2}},
  \href{http://dx.doi.org/10.1016/j.cpc.2015.01.024}{\emph{Comput. Phys.
  Commun.} {\bf 191} (2015) 159--177},
  [\href{http://arxiv.org/abs/1410.3012}{{\tt 1410.3012}}].

\bibitem{deFavereau:2013fsa}
{\scshape DELPHES 3} collaboration, J.~de~Favereau, C.~Delaere, P.~Demin,
  A.~Giammanco, V.~Lema\^\i{}tre, A.~Mertens et~al., \emph{{DELPHES 3, A
  modular framework for fast simulation of a generic collider experiment}},
  \href{http://dx.doi.org/10.1007/JHEP02(2014)057}{\emph{JHEP} {\bf 02} (2014)
  057}, [\href{http://arxiv.org/abs/1307.6346}{{\tt 1307.6346}}].

\bibitem{Goncalves:2015prv}
D.~Goncalves, F.~Krauss and R.~Linten, \emph{{Distinguishing b-quark and gluon
  jets with a tagged b-hadron}},
  \href{http://dx.doi.org/10.1103/PhysRevD.93.053013}{\emph{Phys. Rev. D} {\bf
  93} (2016) 053013}, [\href{http://arxiv.org/abs/1512.05265}{{\tt
  1512.05265}}].

\bibitem{Kim:2019wns}
J.~H. Kim, M.~Kim, K.~Kong, K.~T. Matchev and M.~Park, \emph{{Portraying Double
  Higgs at the Large Hadron Collider}},
  \href{http://dx.doi.org/10.1007/JHEP09(2019)047}{\emph{JHEP} {\bf 09} (2019)
  047}, [\href{http://arxiv.org/abs/1904.08549}{{\tt 1904.08549}}].

\bibitem{Tanabashi:2018oca}
{\scshape Particle Data Group} collaboration, M.~Tanabashi et~al.,
  \emph{{Review of Particle Physics}},
  \href{http://dx.doi.org/10.1103/PhysRevD.98.030001}{\emph{Phys. Rev. D} {\bf
  98} (2018) 030001}.

\bibitem{Cowan:1998ji}
G.~Cowan, \emph{{Statistical data analysis}}.
\newblock 1998.

\end{thebibliography}\endgroup

\bibliographystyle{JHEP}   

\end{document}